\documentclass[aip, preprint, twocolumn, amsmath, amssymb, reprint, superscriptaddress,floatfix]{revtex4-1}

\newcommand{\Sec}[1]{Sec.\,\ref{#1}}

\newcommand{\be}{\begin{equation}}
	\newcommand{\ee}{\end{equation}}
\newcommand{\bea}{\begin{eqnarray}}
	\newcommand{\eea}{\end{eqnarray}}
\newcommand{\Fig}[1]{Fig.\,\ref{#1}}
\newcommand{\Eq}[1]{Eq.\,(\ref{#1})}

\newcommand{\ra}{\rangle}

\newcommand{\RNum}[1]{\uppercase\expandafter{\romannumeral #1\relax}}
\newcommand{\cmark}{\ding{51}}%

\usepackage{graphicx,verbatim}
\usepackage{dcolumn}
\usepackage{bm}
\usepackage{bbm}
\usepackage{sidecap}
\usepackage{braket}
\usepackage{color}
\usepackage{amssymb}
\usepackage{MnSymbol}
\usepackage{csquotes}
\usepackage{pifont}

\usepackage{tikz}
\usepackage{circledsteps}

\newcommand*{\rom}[1]{\expandafter\@slowromancap\romannumeral #1@}

\begin{document}

\title{
Current-induced bond rupture in single-molecule junctions: 
Effects of multiple electronic states and vibrational modes
}

\author{Yaling Ke}
\affiliation{
	Institute of Physics, University of Freiburg, Hermann-Herder-Strasse 3, 79104 Freiburg, Germany
}

\author{Jan Dvo\v{r}\'ak}
\affiliation{Charles University, Faculty of Mathematics and Physics, Institute of Theoretical Physics, V Hole\v{s}ovi\v{c}k\'ach 2, 18000 Prague 8, Czech Republic
}

\author{Martin \v{C}\'i\v{z}ek}
\affiliation{Charles University, Faculty of Mathematics and Physics, Institute of Theoretical Physics, V Hole\v{s}ovi\v{c}k\'ach 2, 18000 Prague 8, Czech Republic
}

\author{Raffaele Borrelli}
\affiliation{DISAFA, University of Torino, Largo Paolo Braccini 2, I-10095 Grugliasco, Italy}

\author{Michael Thoss}
\affiliation{
Institute of Physics, University of Freiburg, Hermann-Herder-Strasse 3, 79104 Freiburg, Germany
}


\begin{abstract}
Current-induced bond rupture is a fundamental process in nanoelectronic architectures such as molecular junctions and in scanning tunneling microscopy measurements of molecules at surfaces. 
The understanding of the underlying mechanisms is important for the design of molecular junctions that are stable at higher bias voltages and is a prerequisite for further developments in the field of current-induced chemistry.
In this work, we analyse the mechanisms of current-induced bond rupture employing a recently developed method, which combines the hierarchical equations of motion approach in twin space with the matrix product state formalism, and allows accurate, fully quantum mechanical simulations of the complex bond rupture dynamics.
Extending previous work [J. Chem. Phys. 154, 234702 (2021)], we consider specifically the effect of multiple electronic states and multiple vibrational modes. The results obtained for a series of models of increasing complexity show the importance of vibronic coupling between different electronic states of the charged molecule, which can enhance the dissociation rate at low bias voltages profoundly. 
 \end{abstract}	
\maketitle

\section{Introduction}\label{introduction}
The prospects of nanoscale electronic devices have been a driving force of the field of molecular electronics.\cite{Aviram_Chem.Phys.Lett._1974_p277--283,Nitzan_Science_2003_p1384--1389,Elbing_Proc.Natl.Acad.Sci.U.S.A._2005_p8815--8820,Cuevas__2010_p,Bergfield_physicastatussolidib_2013_p2249--2266,Aradhya_Nat.Nanotechnol._2013_p399,Baldea__2016_p,Su_Nat.Rev.Mater._2016_p16002,Seideman__2016_p,Thoss_J.Chem.Phys._2018_p030901,Xin_Nat.Rev.Phys._2019_p211--230,Evers_Rev.Mod.Phys._2020_p035001} A typical setup in this field is a single-molecule junction, where a molecule is connected to bulk metal electrodes.
Molecular junctions represent a unique architecture to investigate molecules in a distinct nonequilibrium situation and, in a broader context, to study
basic mechanisms of charge and energy transport in a many-body
quantum system at the nanoscale.

Although the flexible structure of molecules can be utilized to design a great variety of desired functionalities, the strong coupling of molecular vibrations to transport electrons leads to current-induced vibrational heating, which often results in bond rupture and mechanical instability of the junctions, particularly at higher bias voltages.\cite{Persson_Surf.Sci._1997_p45--54,Kim_Phys.Rev.Lett._2002_p126104,Huang_NanoLett._2006_p1240--1244,Schulze_Phys.Rev.Lett._2008_p136801,Ioffe_Nat.Nanotechnol._2008_p727--732,Sabater_BeilsteinJ.Nanotechnol._2015_p2338-2344,Li_J.Am.Chem.Soc._2016_p16159-16164,Capozzi_NanoLett._2016_p3949--3954,Peiris_Chem.Sci._2020_p5246-5256,Bi_J.Am.Chem.Soc._2020_p3384--3391} 
The process of current-induced bond rupture has also been observed experimentally in scanning tunneling microscopy (STM) studies of molecules at surfaces.\cite{stipe1997single,ho2002single,huang2013single}  A comprehensive investigation of the underlying reaction mechanisms of bond rupture is not only crucial for designing molecular junctions that are stable at higher bias voltages,  but is also critical to the development of nano-scale chemical catalysis.\cite{Persson_Surf.Sci._1997_p45--54,Kuznetsov_Electrochem.Commun._2007_p1624--1628,Kolasinski__2012_p,Zhao_Phys.Chem.Chem.Phys._2013_p12428--12441,Dzhioev_J.Chem.Phys._2013_p134103,Cui_Nat.Nanotechnol._2018_p122--127,Kuperman_NanoLett._2020_p5531--5537,Albrecht_Science_2022_p298--301} 

Recently, we have systematically analyzed the basic mechanisms 
of current-induced bond rupture in single-molecule junctions based on a minimal model comprising one electronic state of the charged molecule and a single vibrational reaction mode.\cite{Erpenbeck_Phys.Rev.B_2018_p235452, Erpenbeck_Phys.Rev.B_2020_p195421,Ke_J.Chem.Phys._2021_p234702} 
The results revealed, even for this minimal model, a complex interplay of electronic and vibrational dynamics, resulting in various mechanisms, which govern current-induced bond rupture in different parameter regimes.\cite{Ke_J.Chem.Phys._2021_p234702} 
However, in polyatomic molecules several electronic states and multiple vibrational modes are expected to be involved in the reaction mechanisms.
For example, extensive studies of photoinduced dissociation dynamics and dissociative electron attachment in smaller organic molecules, such as pyrrole and formic acid, in the gas phase have revealed that  electronic states of different character are involved in the reactions and out-of-plane vibrations can mediate the coupling of different electronic states and thus provide an effective dissociation pathway.\cite{Muendel_Phys.Rev.A_1985_p181,Skalicky_Phys.Chem.Chem.Phys._2002_p3583--3590,Rescigno_Phys.Rev.Lett._2006_p213201,Chung_Phys.Chem.Chem.Phys._2007_p2075--2084,Oliveira_J.Chem.Phys._2010_p204301,Janeckova_Phys.Rev.Lett._2013_p213201,Slaughter_Phys.Chem.Chem.Phys._2020_p13893--13902,Modelli_J.Phys.Chem.A_2001_p5836--5841,Vallet_J.Chem.Phys._2005_p144307,Dvorak_Phys.Rev.A_2022_p062821,ragesh2022distant} An important mechanism in this context is photoinduced or electron-induced dissociation involving a $\pi^*\rightarrow \sigma^*$ electronic transition triggered by vibronic coupling.\cite{Modelli_J.Phys.Chem.A_2001_p5836--5841,Vallet_J.Chem.Phys._2005_p144307,Nag_Phys.Rev.A_2021_p032830,Dvorak_Phys.Rev.A_2022_p062821,ragesh2022distant} Little is known about the corresponding reaction mechanisms in the context of molecular junctions, where the molecule is persistently driven out of equilibrium by an electrical current.

In this paper, we address these more complex situations and extend our previous studies of current-induced bond rupture in molecular junctions\cite{Erpenbeck_J.Chem.Phys._2019_p191101,Erpenbeck_Phys.Rev.B_2020_p195421,Ke_J.Chem.Phys._2021_p234702,erpenbeck2022electrical} to models with multiple electronic states and multiple vibrational modes. To tackle this challenging problem, we use the
hierarchical equations of motion (HEOM) method\cite{Tanimura_J.Phys.Soc.Jpn._1989_p101--114,Jin_J.Chem.Phys._2008_p234703,Shi_J.Chem.Phys._2009_p084105,Ye_WIREsComputMolSci_2016_p608--638,Shi_J.Chem.Phys._2018_p174102,Schinabeck_Phys.Rev.B_2020_p075422,Tanimura_J.Chem.Phys._2020_p020901,Baetge_Phys.Rev.B_2021_p235413} in combination with a discrete value representation (DVR)\cite{Colbert_J.Chem.Phys._1992_p1982--1991,Echave_Chem.Phys.Lett._1992_p225--230,Seideman_J.Chem.Phys._1992_p4412-4422} of vibrational modes, as well as the introduction of a dissociation-motivated Lindbladian term with a complex absorbing potential.
This method was introduced before by Erpenbeck et al. \cite{Erpenbeck_Phys.Rev.B_2020_p195421} in the context of current-induced bond rupture in simpler models.
The application to models with multiple electronic states and multiple vibrational modes requires a further extension of the method, because the conventional HEOM approach would require a too large amount of memory resources to store the enormous number of auxiliary density operators (ADOs).
To facilitate the HEOM treatment of these systems, we use an approach developed recently, which maps the HEOM method for a set of ADOs, originally represented in Hilbert space as matrices, into a time-dependent Schr\"odinger-like equation for an extended pure state wavefunction in twin space.\cite{Borrelli_J.Chem.Phys._2019_p234102,Ke_J.Chem.Phys._2022_p194102} For the latter, the well-established matrix product state (MPS) formalism, also called tensor train (TT),\cite{White_Phys.Rev.Lett._1992_p2863,Verstraete_Phys.Rev.Lett._2004_p207204,Oseledets_SIAMJ.Sci.Comput._2011_p2295--2317,Schollwoeck_Ann.Phys.NY_2011_p96--192} and the corresponding tangent-space time propagation schemes\cite{Haegeman_Phys.Rev.B_2013_p075133,Haegeman_Phys.Rev.B_2016_p165116,Paeckel_Ann.Phys.NY_2019_p167998a} can be applied.

The rest of this paper is organized as follows: In \Sec{MM} we introduce the model, outline the method, and provide the definitions of observables. The numerical results of dissociation dynamics and underlying reaction mechanisms are presented and analyzed in \Sec{results}. \Sec{conclusion} concludes with a summary and gives an outlook of future work.

\section{Model and method}\label{MM}
\subsection{Model}\label{Model}
\begin{figure}
	\begin{minipage}[c]{0.45\textwidth} 		
			\raggedright a) \\
	\includegraphics[width=\textwidth]{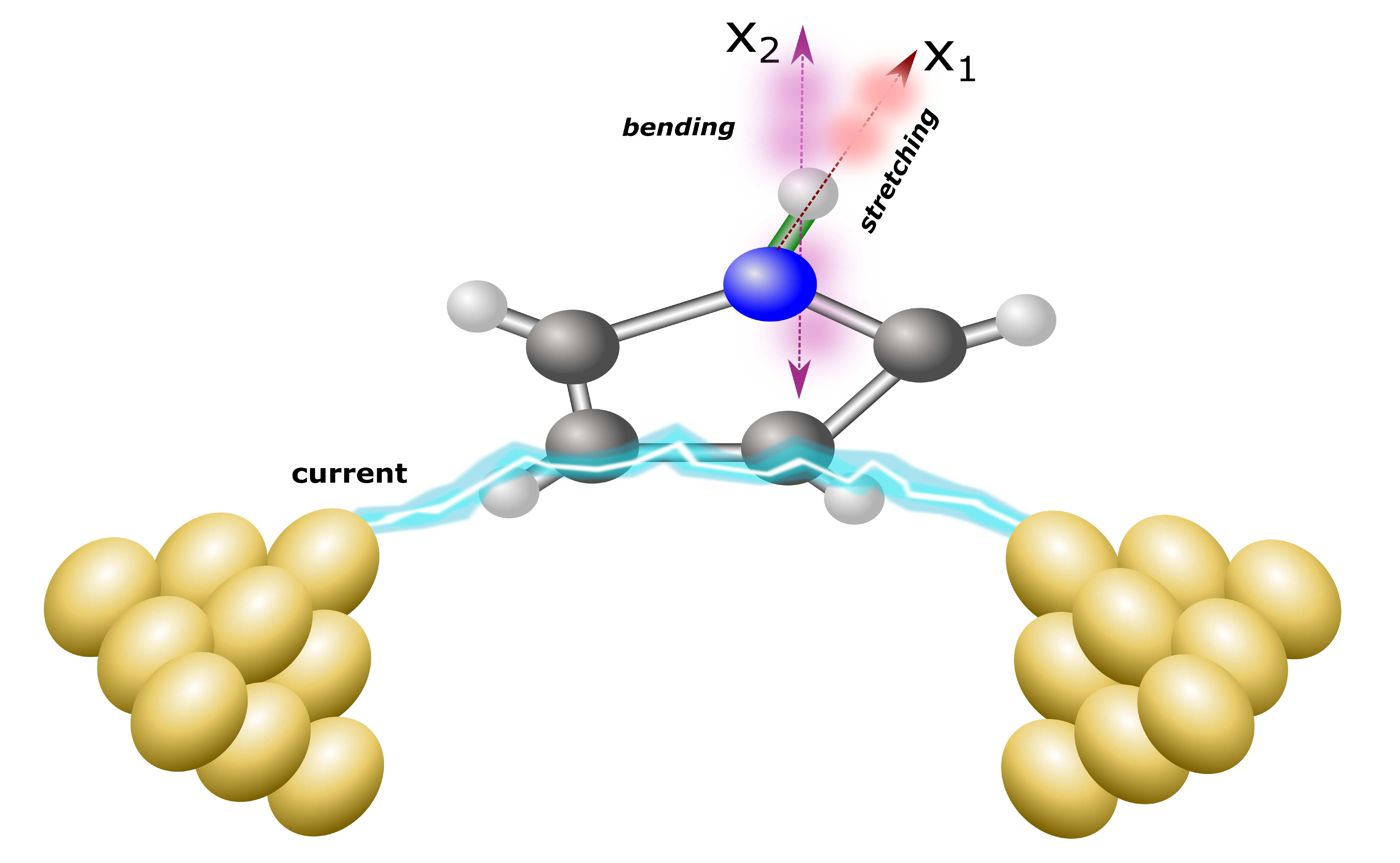}
	\end{minipage}
		\begin{minipage}[c]{0.45\textwidth} 		
			\raggedright b) \\
	\includegraphics[width=\textwidth]{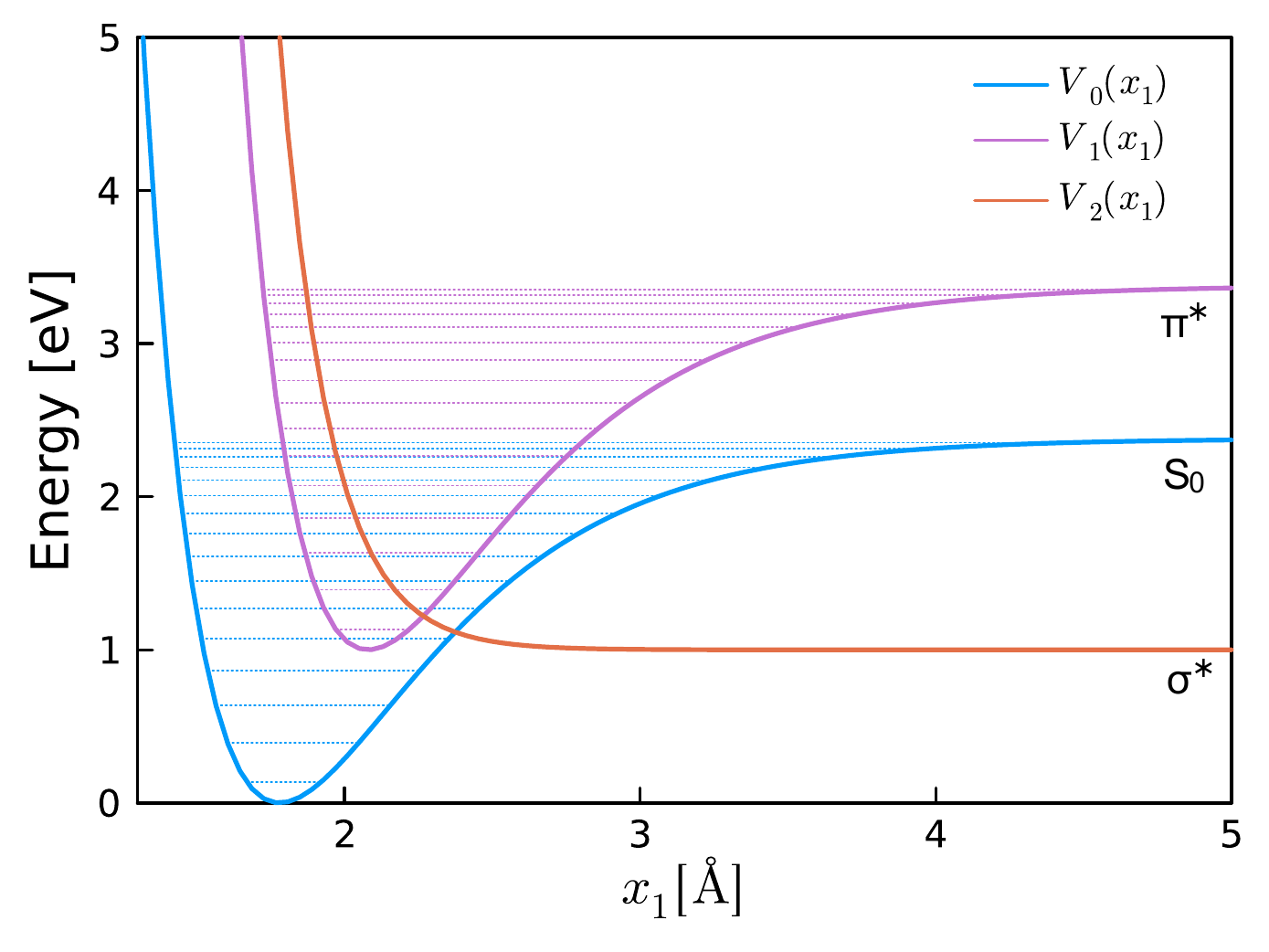}
	\end{minipage}
\caption{(a) Schematic representation of current-induced bond stretching and bending in a molecular junction. (b) Potential energy surfaces along the stretching reaction mode $x_1$ of the electronic ground ($S_0$) state of the neutral molecule  (blue) as well as two diabatic states of the singly charged molecule, corresponding to $\pi^*$ (magenta) and $\sigma^*$ (orange) states. The bending mode is placed at its equilibrium position, i.e. $x_2=0$. The horizontal dotted lines correspond to vibrational energy levels in the respective states.}
\label{energy_profile}	
\end{figure}

In this work, we consider a molecular junction, as depicted in \Fig{energy_profile} (a), where a molecule is connected to two macroscopic leads. The molecule consists of a backbone and a side group. The bond between the side group and backbone can be stretched and it is represented by a reaction coordinate $x_1$. If the bond is elongated beyond a certain length and ruptures, the detachment of the side group occurs. Moreover, the side group can also move out of the backbone plane and this bending coordinate is denoted as $x_2$.  

For illustration, we show in \Fig{energy_profile} (a) as an example the current-induced detachment of a hydrogen atom bonded to the nitrogen atom in a pyrrole molecule. In this class of aromatic molecules, studies in the context of photodissociation and dissociative electron attachment have shown the cooperative effects of multiple electronic states and vibrational modes in the reactions.\cite{Muendel_Phys.Rev.A_1985_p181,Modelli_J.Phys.Chem.A_2001_p5836--5841,Skalicky_Phys.Chem.Chem.Phys._2002_p3583--3590,Vallet_J.Chem.Phys._2005_p144307,Rescigno_Phys.Rev.Lett._2006_p213201,Chung_Phys.Chem.Chem.Phys._2007_p2075--2084,Oliveira_J.Chem.Phys._2010_p204301,Janeckova_Phys.Rev.Lett._2013_p213201,Slaughter_Phys.Chem.Chem.Phys._2020_p13893--13902,Dvorak_Phys.Rev.A_2022_p062821,ragesh2022distant} In particular, a dissociation mechanism involving the vibronic coupling between $\pi^*$ (magenta) and $\sigma^*$ (orange) states has been found to be of importance. 
In molecular junctions under finite bias voltage, as considered here, the situation is more complex, because the electrical current through the molecule results in a genuine nonequilibrium situation which allows other reaction mechanisms.


We use a generic model of a molecular junction, given by 
the system-bath Hamiltonian 
\begin{equation}
\label{total_hamiltonian}
	H=H_{\rm S}+H_{\rm B}+H_{\rm SB}.
\end{equation}
Here, the system Hamiltonian $H_{\rm S}$ describes the molecule, the bath Hamiltonian $H_{\rm B}$ models the macroscopic electrodes and $H_{\rm SB}$ is the molecule-electrode coupling.

For the molecule, a model is adopted, where two electronic states of the charged molecule are taken into account as well as two vibrational modes as highlighted in \Fig{energy_profile} (a). Correspondingly, the system Hamiltonian is expressed as (we set $e=\hbar=k_B=1$)
\begin{equation}
\label{systemHamiltonian}
\begin{split}
	H_{\rm S}=&T_{\rm nuc} +V_0(x_1,x_2)+ \epsilon_1(x_1) d_1^+d_1^- + \epsilon_2(x_1) d_2^+d_2^-  \\
	&+Ud_1^+d_1^-d_2^+d_2^-
	 +\Delta(x_2)\left(d_1^+d_2^-+d_2^+d_1^-\right).
\end{split}
\end{equation}
Here, $T_{\text{nuc}}$ is the nuclear kinetic energy operator and $V_0(x_1,x_2)$ denotes the potential energy surface (PES) of the electronic ground state of the neutral molecule (labeled as $S_0$ in what follows), which is spanned along the two vibrational modes $x_1$ and $x_2$. We assume that $V_0(x_1,x_2)$ is described by a Morse function along the stretching mode $x_1$, and the bending mode $x_2$ is characterized for simplicity as a harmonic oscillator,
\begin{equation}
V_0(x_1,x_2) =D_e(1-e^{-a(x_1-x_1^0)})^2  +  \frac{1}{2}\omega_b x_2^2.
\end{equation}
In the calculations reported below, we have chosen representative parameters similar as in our previous studies\cite{Erpenbeck_J.Chem.Phys._2019_p191101,Erpenbeck_Phys.Rev.B_2020_p195421,Ke_J.Chem.Phys._2021_p234702}: mass of stretching mode $m_s=1$ amu, dissociation energy $D_e = 2.38$ eV, width parameter of the Morse potential $a=1.028 \text{ \AA}^{-1}$, the equilibrium distance $x_1^0 = 1.78 \text{ \AA}$. 
For the bending mode,  the harmonic frequency $\omega_b=145$ meV is adopted, and the dimensionless coordinate is expressed as $x_2=(b^++b^-)/\sqrt{2}$, where $b^+$ and $b^-$ denote the corresponding creation and annihilation operators, respectively. Although our model is inspired by the pyrrole molecule, we emphasize
that the goal of this work is to study the basic mechanisms of current-induced bond rupture and we do not attempt to describe a specific molecule.

The operators $d_{i}^{+}/d_{i}^-$ in \Eq{systemHamiltonian} are linked to the creation/annihilation of an electron in the $i$th electronic state, and $\epsilon_i(x_1)$ is the charging energy of the corresponding electronic state at a fixed point $x_1$. Thus, the PESs of two singly charged anionic states (assigned with the notation $\pi^*$ and $\sigma^*$, respectively) are obtained as $V_i(x_1,x_2)=V_0(x_1, x_2)+\epsilon_i(x_1)$. We assume that the PESs of two electronic states, according to there different electronic character ($\pi^*$ and $\sigma^*$), have distinctively different characteristics and are modeled by 
\be
V_1(x_1,x_2=0)=D_e (1-e^{-a(x_1-x_1')})^2 + E_1,
\ee
\be
V_2(x_1, x_2=0)=A_e e^{-a'(x_1-x_1^0)} + E_2.
\ee
In the planar geometry, i.e. $x_2=0$, the PES of the $\pi^*$ charged state has the same profile as that of the neutral $S_0$ state, but the equilibrium position is displaced to a larger bond distance $x_1'=2.08\text{ \AA}$ along with a shift in the energy of $E_1=1$ eV. The other charged state $\sigma^*$ has an anti-bonding character and is modeled by a repulsive exponential function with the following parameters: $A_e=4$ eV, $a'=5.958 {\text{ \AA}^{-1}}$ and $E_2=1$ eV. All the PESs are displayed in \Fig{energy_profile} (b) and, as can be seen therein, the PES of the $\sigma^*$ state intersects with those of both $S_0$ and $\pi^*$ states.

Coulomb electron-electron interaction is quantified by the parameter $U$ and the coupling between two diabatic states by $\Delta(x_2)$. We assume that the bending mode is nonreactive but could mediate the diabatic coupling between two electronic charged states, and the coupling takes the form
\begin{equation}
\label{diabatic_term}
   \Delta(x_2) = \Delta_0+\Delta_1x_2e^{-\lambda x_2^2}.
\end{equation}
In the calculations reported below, the coupling parameters are chosen as $\Delta_0=\Delta_1=0.5$ eV. A more detailed discussion of the diabatic coupling and the role of the two coupling parameters $\Delta_0, \Delta_1$ is given in the SI.

The electrodes are modeled as noninteracting electron reservoirs
\begin{equation}
H_B= \sum_{\alpha}H_B^{\alpha} = \sum_{\alpha k}\epsilon_{\alpha k}c^{+}_{\alpha k}c_{\alpha k}^-
\end{equation}
where $c^{+}_{\alpha k}/c_{\alpha k}^-$ denotes the creation/annihilation operator of electronic state $k$ in lead $\alpha$ associated with the energy $\epsilon_{\alpha k}$. 

If an external voltage bias $\Phi$ is applied upon the junction, electrons can be transferred from electrodes to the molecular bridge or vice versa. Their coupling term is described by the Hamiltonian
\begin{equation}
    H_{SB} = \sum_{i} \sum_{\alpha k}v_{i \alpha k}(x_1, x_2)c^{+}_{\alpha k}d_i^- + h.c.
\end{equation}

Given the above linear form of the coupling, the influence of electronic reservoirs on the dynamics of the molecule can be characterized completely by the correlation function
\begin{equation}
\label{bath_correlation_function}
    C^{\sigma}_{i \alpha}(t, x_1, x_2) = \int e^{i\sigma \epsilon t}\Gamma_{i\alpha}(\epsilon, x_1,x_2)f^{\sigma}(\epsilon) d\epsilon. 
\end{equation}
The spectral density function $\Gamma_{i\alpha}(\epsilon, x_1,x_2)$ is given by 
\begin{equation}
\label{spectraldensityfunction}
    \Gamma_{i\alpha}(\epsilon, x_1,x_2)=2\pi \sum_k |v_{i \alpha k}(x_1,x_2)|^2 \delta(\epsilon-\epsilon_{\alpha k}),
\end{equation}
which encodes the information of the density of states in lead $\alpha$ as well as the interaction between the $i$th molecular electronic state and all electronic states in lead $\alpha$ at a given nuclear configuration $(x_1, x_2)$. For the sake of simplicity, in this work, we adopt the wide-band approximation and assume that $\Gamma_{i\alpha}(\epsilon, x_1,x_2)$ is a coordinate-independent constant value $\Gamma=\Gamma_{i\alpha}=v_{i\alpha}^2$. This quantity also determines the timescale of electron transfer between the electrodes and the central molecule. However, we should mention that the generalization to a structured environment and a coordinate-dependent molecule-lead coupling is in principle straightforward.\cite{erpenbeck2022electrical} 
The electron distribution in lead $\alpha$ in equilibrium is represented by the Fermi function
\begin{equation}
   f^{\sigma}_{\alpha}(\epsilon)= \frac{1}{1 + e^{\sigma (\epsilon - \mu_{\alpha})/T}}.
\end{equation}
Here, 
$T$ is the temperature, $\mu_{\alpha}$ the chemical potential, and $\sigma=\pm$. 

\subsection{Method}
To study the current-induced bond rupture dynamics in a single-molecule junction model described in \Sec{Model}, we use the HEOM method. This numerically exact hierarchical quantum master equation approach generalizes perturbative quantum master equation methods by including higher-order contributions as well as non-Markovian memory and allows for the systematic convergence of the results. For more details about the developments of the HEOM method, we refer to the review in Ref.\ \onlinecite{Tanimura_J.Chem.Phys._2020_p020901} and the references therein. The development of the HEOM method for simulations of vibrationally coupled electron transport in molecular junctions as well as current-induced bond rupture is described in Refs.\ 
\onlinecite{schinabeck2016hierarchical,schinabeck2018hierarchical,Schinabeck_Phys.Rev.B_2020_p075422,Erpenbeck_J.Chem.Phys._2019_p191101,Erpenbeck_Phys.Rev.B_2020_p195421,Ke_J.Chem.Phys._2021_p234702}.

A core idea underlying the HEOM method is to expand the correlation function in \Eq{bath_correlation_function} as a sum of exponential functions, by virtue of sum-over-pole decomposition schemes of the Fermi distribution function,\cite{Hu_J.Chem.Phys._2010_p101106,Hu_2011_J.Chem.Phys._p244106} 
\begin{equation}
\label{bath_correlation_function_exponentials}
    C^{\sigma}_{i \alpha}(t) \simeq v_{i\alpha}^2\delta(t)/2+\sum_{p=1}^P v_{i\alpha}^2\eta_{i\alpha p} e^{-i\gamma_{i\alpha p}^{\sigma} t}. 
\end{equation}
The equation holds exactly when the number of poles $P\rightarrow \infty$. However, at finite temperatures, a finite $P$ is usually adequate to well reproduce the original correlation function. We adopt here the Pad\'e pole decomposition scheme and the explicit expression of $\eta_{i\alpha p}$ and $\gamma_{i\alpha p}^{\sigma}$ can be found in Refs. \onlinecite{Hu_J.Chem.Phys._2010_p101106,Hu_2011_J.Chem.Phys._p244106}, but other choices suitable for lower temperatures are possible.\cite{Zhang_J.Chem.Phys._2020_p064107,chen2022universal,xu2022taming}

As discussed in more detail below, the exponential expansion in \Eq{bath_correlation_function_exponentials} can be interpreted as a mapping of the continuous infinite set of electronic degrees of freedom of the electrodes electrons into an effective fermionic environment with a finite number of virtual discrete electronic levels. In this effective fermionic bath, there are in total $K=2N_eN_{\alpha}P$ virtual levels and each is specified by four indices, $(i, \alpha, p, \sigma)$, i.e. the electronic index $i \in \{1,\cdots, N_e\}$, the lead index $\alpha \in \{1,\cdots, N_{\alpha}\}$, the pole index $p \in \{1,\cdots, P\}$ linked to the decomposition in \Eq{bath_correlation_function_exponentials}, and the sign index $\sigma= \pm \,(\bar{\sigma}=-\sigma)$. The occupancy of the $k$th virtual level is denoted by $n_k$ (empty when $n_k=0$ and filled when $n_k=1$). 

For each configuration of the ordered set $\bm{n}=(n_1, n_2,\cdots, n_K)$, an auxiliary density operator (ADO) $\rho^{\bm{n}}(t)$ can be introduced. In particular, the reduced system dynamics is reproduced by the zeroth order ADO, $\rho_S=\rho^{(0, 0,\cdots, 0)}$, where all these virtual electronic levels are unpopulated.  The joint system-bath dynamics is encoded into higher order ADOs, which altogether can be obtained by propagating the following hierarchical set of equations of motion,
\begin{widetext}
\begin{eqnarray}
\label{heom}
\frac{d \rho^{\bm{n}}(t) }{dt} 
&=&- i\left[H_{\rm{S}},\rho^{\bm{n}}(t)\right]_- -\mathcal{L}^{\infty} \rho^{\bm{n}}(t)
+{\sum_{k=1}^K n_{k}\gamma^{\sigma_k}_{i_k\alpha_k p_k}  \rho^{\bm{n} }  }(t) 
-\sum_{i\alpha \sigma} \frac{v^2_{i\alpha}}{4}
\left[d_{i}^{\bar{\sigma}}, \left[d_{i}^{\sigma}, \rho^{\bm{n}}(t)\right]_{(-)^{||\bm{n}||+1}}\right]_{(-)^{||\bm{n}||+1}}\\
 &&+i\sum_{k=1}^{K} (-1)^{\sum_{j<k}n_{j}}  \sqrt{1-n_k} v_{i_k \alpha_k}\left(  d_{i_k}^{\bar{\sigma}_k} \rho^{\bm{n}+{\bm{1}}_k}(t)+
   (-1)^{||\bm{n}||+1}\rho^{\bm{n}+\bm{1}_k} (t)d^{\bar{\sigma}_k}_{i_k} \right)\nonumber \\
&& +i\sum_{k=1}^{K} (-1)^{\sum_{j<k}n_{j}}  \sqrt{n_k} v_{i_k \alpha_k}\left(\eta_{i_k\alpha_k p_k} d^{\sigma_k}_{i_k} \rho^{\bm{n}-\bm{1}_k}(t)  
{ -(-1)^{||\bm{n}||-1}}
 \rho^{\bm{n}-\bm{1}_k}(t)  \eta_{i_k\alpha_kp_k}^{*} d^{\sigma_k}_{i_k}\right). \nonumber 
\end{eqnarray}
\end{widetext}
Here, 
$[\mathcal{O}, \rho^{\bm{n}}]_-$ and $[\mathcal{O}, \rho^{\bm{n}}]_+$ denote the commutator and anticommutator of an operator $\mathcal{O}$ and ADO $\rho^{\bm{n}}$, respectively. The notation $\bm{n}\pm \bm{1}_k$ is given by
\begin{equation}
\bm{n}\pm \bm{1}_k = (n_1, n_2, \cdots, 1-n_k, \cdots, n_K).
\end{equation}

To describe the vibrational dynamics of the dissociative reaction mode $x_1$, a sine-DVR representation is employed.\cite{Colbert_J.Chem.Phys._1992_p1982--1991} Specifically, $x_1$ is represented in a  range from $x_1^{\rm min}=1.35\text{ \AA}$ to $x_1^{\rm max}=5\text{ \AA}$ with $N_{\rm DVR}$ grid points.



Furthermore, in order to avoid finite size effects, we introduce in \Eq{heom} a physically motivated Lindblad term,\cite{Erpenbeck_Phys.Rev.B_2020_p195421}
\begin{equation}
\label{Lindblad}
\begin{split}
\mathcal{L}^{\infty} \rho^{\bm{n}} = \sum_{j=1}^{N_{\rm DVR}}&W(x^j_1)\left(|x^{j}_1\rangle \langle x^j_1|\rho^{\bm{n}} +\rho^{\bm{n}}|x^{j}_1\rangle \langle x^j_1| \right) \\
&-2W(x^j_1)|x^{\infty}_1\rangle \langle x^j_1| \rho^{\bm{n}}|x^j_1\rangle \langle x^{\infty}_1|,
\end{split}
\end{equation}
which absorbs the vibrational wave packet from DVR grid points $x_1^j$ in the finite-size region onto  an additional grid point $x_1^{\infty}$, which is representative of large distances $x_1$ of the detached side group. This is  achieved by the complex absorbing potential (CAP),
\begin{equation}
\label{CAP}
   W(x_1) = ig \cdot (x_1 - x_1^{\rm CAP})^4 \cdot \Theta(x_1 - x_1^{\rm CAP}),  
\end{equation}
where $g=5 \text{ eV/}\text{\AA}^4$, and $\Theta$ denotes the Heaviside step function, i.e. absorption of the wave packet is only activated beyond a certain bond length, which in the calculations reported below is chosen as $x_1^{\mathrm CAP}=4.0\text{ \AA}$. 
The second term on the right-hand side of \Eq{Lindblad} compensates for the loss of the norm of the density matrix introduced by the CAP.
The parameters of the CAP were determined by test calculations to ensure that the observables obtained do not depend on the CAP.


Employing \Eq{heom} to obtain current-induced dissociation dynamics in single-molecule junctions is in principle straightforward, but it quickly becomes infeasible when multiple electronic states and vibrational modes are taken into account, because it requires a large amount of computational memory. To circumvent this problem, one can reformulate \Eq{heom} into a Schr\"odinger-like equation, which facilitates the application of MPS/TT decomposition schemes.

To this end, instead of representing an ADO as a density matrix in Hilbert space, it is recast into a rank-$2D$ tensor in the so-called twin space with $D$ being the number of system degrees of freedom (DoFs),\cite{Schmutz_ZeitschriftfurPhysikBCondensedMatter_1978_p97--106,Suzuki_J.Phys.Soc.Jpn._1985_p44834485,Arimitsu_Prog.Theor.Phys._1987_p32--52,Feiguin_Phys.Rev.B_2005_p220401,Borrelli_WIREsComputMolSci_2021_pe1539}
\begin{equation}
    |\rho^{\bm{n}}(t)\rrangle =\sum_{s_1\tilde{s}_1\cdots s_D\tilde{s}_D} C^{\bm{n}}_{s_1\tilde{s}_1\cdots  s_D\tilde{s}_D}(t)|s_1\tilde{s}_1\cdots  s_D\tilde{s}_D\rangle.
\end{equation}
Besides, for every single-site operator in \Eq{systemHamiltonian}, there is a pair of counterpart super-operators in twin space, $\hat{d}^{\pm}_i$ and $\tilde{d}^{\pm}_i$, as well as $\hat{x}_j$ and $\tilde{x}_j$, acting on $|\rho\rrangle$ as
\begin{subequations}
\begin{align}
     \hat{d}^{\pm}_i  |\rho\rrangle &= d^{\pm}_i \otimes \mathbbm{1}_i^e |\rho\rrangle \coloneq d^{\pm}_i \rho ,\\
    \tilde{d}^{\pm}_i  |\rho\rrangle &= \mathbbm{1}_i^e \otimes d^{\mp}_i |\rho\rrangle \coloneq \rho d^{\mp}_i      ,\\
    \hat{x}_j  |\rho\rrangle &= x_j \otimes \mathbbm{1}_j^{ \rm{vib}} |\rho\rrangle \coloneq x_j \rho ,\\
    \tilde{x}_j  |\rho\rrangle &= \mathbbm{1}_j^{\rm{vib} } \otimes x_j |\rho\rrangle \coloneq \rho x_j .
\end{align}
\end{subequations}
The super-operators with a hat (\enquote{$\,\, \hat{}\,\,$}) act on the physical DoFs, while those with a tilde (\enquote{$\,\,\tilde{}\,\,$}) act on ancilla DoFs. For more theoretical and technical details with regard to this transformation, we refer the reader to Refs. \onlinecite{Schmutz_ZeitschriftfurPhysikBCondensedMatter_1978_p97--106,Suzuki_J.Phys.Soc.Jpn._1985_p44834485,Arimitsu_Prog.Theor.Phys._1987_p32--52,Feiguin_Phys.Rev.B_2005_p220401,Borrelli_WIREsComputMolSci_2021_pe1539,Ke_J.Chem.Phys._2022_p194102}.

Furthermore, as implied before, $n_k$ denotes the occupation number of the virtual effective electronic level $k$ in the leads. Generating or annihilating an electron at this level is introduced by acting a pair of ad-hoc creation and annihilation operators, $c_k^{<,+}$ (or $c_k^{>,+}$) and $c_k^{>,-}$ (or $c_k^{>,-}$) on the Fock state $|\bm{n}\rangle=|n_1 n_2\cdots n_K\rangle$,
\begin{subequations}
\begin{align}
c^{\lessgtr,+}_{k}& |\bm{n}\rangle=(-1)^{\sum_{j\lessgtr k}n_j}\sqrt{1-n_k}|\bm{n}+\bm{1}_k\rangle,\\
c^{\lessgtr,-}_{k}& |\bm{n}\rangle=(-1)^{\sum_{j\lessgtr k}n_j}\sqrt{n_k}|\bm{n}-\bm{1}_k\rangle,\\
c^{<,+}_{k}& c^{<,-}_{k}|\bm{n}\rangle=n_k|\bm{n}\rangle,\\
 I^{>}&|\bm{n}\rangle=(-1)^{\sum_{j=1}^{K}n_j}|\bm{n}\rangle.
 \end{align}
\end{subequations}
Using the Jordan-Wigner transformation,\cite{jordan1993paulische,Nielsen_SchoolofPhysicalSciencesTheUniversityofQueensland_2005_p} these operators can be represented explicitly in terms of spin operators as 
\begin{equation}
\begin{split}
    c_k^{<, +} \mapsto \left(\bigotimes_{l=1}^{k-1} \sigma_l^z\right)\otimes \sigma_k^+ &\text{ and }
    c_k^{>, +} \mapsto \sigma_k^+ \otimes \left(\bigotimes_{l=k+1}^{K} \sigma_l^z\right)\\
        c_k^{<,-} \mapsto \left(\bigotimes_{l=1}^{k-1} \sigma_l^z\right)\otimes \sigma_k^- &\text{ and }
    c_k^{>,-} \mapsto \sigma_k^- \otimes \left(\bigotimes_{l=k+1}^{K} \sigma_l^z\right),
    \end{split}
\end{equation}
where
\begin{equation}
    \sigma_k^+ = \left(\begin{array}{ll} 0 & 1\\ 0 & 0\end{array}\right), \quad
    \sigma_k^- = \left(\begin{array}{ll} 0 & 0\\ 1 & 0\end{array}\right), \quad
    \sigma_k^z = \left(\begin{array}{ll} 1 & 0\\ 0 & -1\end{array}\right)
\end{equation}
are $2\times 2$ spin matrices. In addition, we have
\begin{equation}
I^{>} \mapsto \bigotimes_{l=1}^{K} \sigma_l^z.
\end{equation}

All the ADOs combined constitute an extended pure state wavefunction in the enlarged space
\begin{equation}
    |\Psi(t)\rangle=\sum_{\begin{subarray}{c} n_1\cdots n_K   \\
   s_1\tilde{s}_1\cdots  s_D\tilde{s}_D\\  \end{subarray}}
    C^{n_1\cdots n_K}_{s_1\tilde{s}_1\cdots  s_D\tilde{s}_D}(t) |n_1\cdots n_K\rangle  |s_1\tilde{s}_1\cdots  s_D\tilde{s}_D\rangle,
\end{equation}
whose time-derivative yields a Schr\"odinger-like equation
\begin{equation}
\label{schroedinger_equation}
i\frac{d |\Psi(t)\rangle }{dt} =\mathcal{H}
 |\Psi(t)\rangle.
\end{equation}
The super Hamiltonian $\mathcal{H}$ in this further enlarged space is explicitly written as
\begin{eqnarray}
\label{Hamiltonian}
\mathcal{H} &=&\hat{H}_S -\tilde{H}_S - \mathcal{L}_{TS}^{\infty}
-i\sum_{k=1}^{K} \gamma^{\sigma_k}_{i_k \alpha_k p_k} c^{<,+}_{k}   c^{<,-}_{k}  \\
&&-\sum_{i \alpha \sigma} \frac{iv^2_{i\alpha}}{4} 
(\hat{d}_{i}^{\bar{\sigma}} - I^{>} \tilde{d}_{i}^{\bar{\sigma}})\cdot (\hat{d}_{i}^{\sigma}-I^{>}\tilde{d}_{i}^{\sigma})  \nonumber\\
&&
-\sum_{k=1}^{K}  v_{i_k \alpha_k} \left( c^{<,-}_{k} \hat{d}^{\bar{\sigma}_k}_{i_k} -  c^{>,-}_{k} \tilde{d}_{i_k}^{\bar{\sigma}_k}\right)  \nonumber\\
 && 
-\sum_{k=1}^{K}v_{i_k \alpha_k} \left( \eta_{i_k\alpha_k p_k}  c^{<,+}_{k}  \hat{d}^{\sigma_k}_{i_k}  -\eta_{i_k\alpha_kp_k}^{*}  c^{>,+}_{k}  \tilde{d}^{\sigma_k}_{i_k}\right), \nonumber
\end{eqnarray}
where $\mathcal{L}^{\infty}_{TS} $ is the corresponding Lindblad operator (\Eq{Lindblad}) in twin space
\begin{equation}
\label{Lindblad_twinspace}
\begin{split}
\mathcal{L}^{\infty}_{TS}  = \sum_{j=1}^{N_{\rm DVR}}&W(x^j_1)\left(|x^{j}_1\rangle \langle x^j_1| +|\tilde{x}^{j}_1\rangle \langle \tilde{x}^j_1| \right) \\
&-2W(x^j_1)|x^{\infty}_1\tilde{x}^j_1\rangle \langle x^j_1\tilde{x}^{\infty}_1|.
\end{split}
\end{equation}

One efficient algorithm to solve \Eq{schroedinger_equation} is to bring $|\Psi\rangle$ into the matrix product state format.  The idea is to decompose the time-dependent  high-rank coefficient tensor $C_{ s_1\tilde{s}_1\cdots s_D \tilde{s}_D}^{n_1\cdots n_K}$ into a product of low-rank matrices, 

\begin{eqnarray}
\label{coefficient_mps}
      C_{ s_1 \tilde{s}_1 \cdots s_D \tilde{s}_D}^{n_1 \cdots n_K}
               &=& \sum_{r_0 r_1  \cdots r_{K+2D}}    A^{[1]}(r_0,n_1,r_1) A^{[2]}(r_1,n_2,r_2)   \cdots \nonumber  \\ && \qquad \quad A^{[K+2D]}(r_{K+2D-1},\tilde{s}_D, r_{K+2D}). 
\end{eqnarray}
The rank-3 tensors $A^{[i]}$ are called the cores of the MPS/TT decomposition. For the physically relevant indices $n_i$ (or $s_i$, $\tilde{s}_i$), $A^{[i]}(n_i)$ is an $r_{i-1}\times r_i$ complex-valued matrix. The dimensions $r_i$ are called compression ranks or bond dimensions. Specifically, the first and the last rank are fixed as  $r_0=r_{K+2D}=1$, such that the matrices multiply into a scalar for a given configuration $|n_1\cdots n_Ks_1\tilde{s}_1s_D \tilde{s}_D\ra$. The decomposition in \Eq{coefficient_mps} is formally exact in the limit of infinite bond dimension, but in practical implementation, a truncation is always needed with a maximally allowed bond dimension $r_{\rm max}$. The numerically exact observables are obtained when the results are converged with respect to $r_{\rm max}$.

In analogy to the MPS description of the wave function, the super Hamiltonian $\mathcal{H}$ can also be efficiently parameterized in the matrix product operator (MPO) format as
\begin{eqnarray}
    \mathcal{H}&=&
        X^{[1]}(n_1,n'_1) \cdots  X^{[K]}(n_K,n'_K) \nonumber \\
        &&X^{[K+1]}(s_1, s'_1)\cdots  X^{[K+2D]}(\tilde{s}_D, \tilde{s}'_D), 
\end{eqnarray}
where $X^{[i]}$ are rank-4 tensors and obtained by repeatedly performing a sequence of Kronecker products, standard MPO addition and single value decomposition (SVD) truncation with a prescribed accuracy $\varepsilon$ to control the ranks of tensor train matrices.\cite{Schollwoeck_Ann.Phys.NY_2011_p96--192}

We employ the one-site version of the time-dependent variational principle (TDVP) scheme,\cite{Haegeman_Phys.Rev.B_2013_p075133,Haegeman_Phys.Rev.B_2016_p165116,Paeckel_Ann.Phys.NY_2019_p167998a} which is well suited to Hamiltonians with long-range coupling in the MPO format.  The method solves the dynamical equations projected onto a manifold $\mathcal{M}$, which is the set of MPS with fixed ranks. The resulting equation of motion is written formally as
\begin{equation}
\label{TDVP}
    \frac{d}{dt}|\Psi(A(t))\rangle = -i P_{T(A(t))}\mathcal{H} | \Psi(A(t))\rangle, 
\end{equation}
where $A$ labels all the cores of the MPS/TT representation. The notation $P_{T(A(t))}$ denotes the orthogonal projection into the tangent space of $\mathcal{M}$ at $| \Psi(A(t))\rangle$. \Eq{TDVP} is solved using a Trotter-Suzuki decomposition of the projector and the solution is the best approximation within the manifold $\mathcal{M}$ to the actual wave function. For a detailed account of the time propagation method in the tangent space we refer to Refs. \onlinecite{Haegeman_Phys.Rev.B_2013_p075133,Haegeman_Phys.Rev.B_2016_p165116,Paeckel_Ann.Phys.NY_2019_p167998a}. 

While in the conventional HEOM method where a hierarchy truncation is indispensable in the practical implementation and the method truncated at a hierarchical depth $L$ is roughly equivalent to a $2L$-order quantum master equation, we should emphasize that all higher-order effects are inherently accounted for in the HEOM+MPS/TT method, as all the ADOs are included in the extended wave function $|\Psi(t)\rangle$. 

\subsection{Observables of interest}

Any system or bath-related observable can be obtained directly from the extended wavefunction $\Psi(t)$, and there is a one-to-one correspondence between each ADO and a reduced state of the extended wave function. For instance, the reduced density operator of the system, $ \rho_S$, is extracted by contracting the environmental sites out with a projection onto $\bm{n}=\bm{0}$, i.e., 
\begin{equation}
    \rho_S=\rho^{\bm{0}} \equiv \langle  \bm{n}=\bm{0} |\Psi(t)\rangle.
\end{equation}
Similarly, a first-tier ADO $\rho^{\bm{1}}_{\bm{a}}$ assigned with a specified superindex $\bm{a}=(i, \alpha, p, \sigma)$ is obtained as
\begin{equation}
   \rho^{\bm{1}}_{\bm{a}} \equiv \langle  \bm{n}=\bm{1}_k |\Psi(t)\rangle,
\end{equation}
where $i_k=i, \alpha_k=\alpha, p_k=p$ and $\sigma_k=\sigma$.

In this work, we are particularly interested in the current-induced dissociation dynamics as well as the general dynamics of the electronic and vibrational degrees of freedom. The latter are characterized by the time-dependent populations of the electronic and vibrational states. Specifically, $\langle s_1s_2x_1^j|\rho_s|x_1^js_2s_1\rangle$ describes the joint probability (density) to find the system at a DVR grid point $x_1^j$ and in the electronic configuration $|s_1s_2\rangle$, Here, $s_1$ specifies the occupancy in the first electronic state that is related to the $\pi^*$ state and $s_2$ for the second state that is related to the $\sigma^*$ state. The level is empty when $s_{1/2}=0$ or filled when $s_{1/2}=1$. This observable at time $t$ is obtained as the expectation value
\begin{equation}
\label{population_wavepacket}
\begin{split}
    P_{s_1s_2}(x_1^j,t) &= \text{Tr}_s\{(d_1^+d_1^-)^{s_1}(d_2^+d_2^-)^{s_2}  \rho^{(0)}(t)|x_1^j\rangle \langle x_1^j| \}\\
    &=\sum_{\begin{subarray}{c} x_2\\ \tilde{s}_{1/2}=s_{1/2}\\ \tilde{x}_{1/2}=x_{1/2}\end{subarray}}\langle \bm{n}=\bm{0} |\langle s_1\tilde{s}_1 s_2\tilde{s}_2 x_2 \tilde{x}_2x_1^j \tilde{x}_1^j|\Psi(t)\rangle,
\end{split}
\end{equation}
where $\text{Tr}_s$ denotes the trace over electronic and vibrational DoFs of the molecular
system, and the tilde indices in the twin-space formulation are identical to their corresponding physical indices, i.e. $\tilde{s}_{1/2}=s_{1/2}$ and $\tilde{x}_{1/2}=x_{1/2}$.

The population of the electronic states is obtained by summing $P_{s_1s_2}(x_s^j)$ over all DVR grid points in the finite bond length region and at the infinity point $x_1^{\infty}$, i.e.
\begin{equation}
 P_{s_1s_2}(t) =\sum_{j=1}^{N_{\rm DVR}}P_{s_1s_2}(x_1^j,t)+P_{s_1s_2}(x_1^{\infty},t).
\end{equation}
For simplicity, we use $P_{S_0}$, $P_{\pi^*}$ and $P_{\sigma^*}$ to denote the population probability in the neutral $S_0$ state (correponding to the completely unoccupied state $P_{00}$), and the two singly charged states $P_{10}$ and $P_{01}$, respectively. To avoid the double occupancy, the population in the dianionic state $P_{11}$ is completely suppressed by assuming a large enough Coulomb interaction $U$.

The dissociation probability is defined as the population at the point $x_1^{\infty}$,
\begin{equation}
    P(x_1^{\infty}, t) = \sum_{s_1s_2}  P_{s_1s_2} (x_1^{\infty}, t). 
\end{equation}

Assuming exponential kinetics of the dissociation process in the long-time limit, the dissociation rate is evaluated as 
\begin{equation}
\label{rate}
k_{\mathrm{diss}}=-\lim_{t\rightarrow \infty}
\frac{d}{dt}\ln(1-P(x_1^{\infty},t)).
\end{equation}

\subsection{Numerical details}
In this section, we provide some details of the numerical calculations presented below.
In the simulations, we assume that the molecule and leads are initially disentangled $\rho(0)=\rho_S(0)\otimes \rho_B(0)$. The initial state of the molecule is given by
\begin{equation}
\rho_S(0)=d_1^-d_1^+\otimes d_2^-d_2^+\otimes |\psi_1^0\rangle \langle \psi_1^0|\otimes |\psi_2^0\rangle \langle \psi_2^0|
\end{equation}
corresponding to the $S_0$ electronic state of the neutral molecule and the associated vibrational ground states $|\psi_1^0\rangle$ and $|\psi_2^0\rangle$ of the two vibrational modes, respectively.
The electrodes are initially described by their grand canonical distribution,
\begin{equation}
\label{initial_rho_leads}
\rho_{B}(0)=\frac{e^{-\sum_{\alpha}\beta_{\alpha}(H_B^{\alpha}-\mu_{\alpha}N_{\alpha})}}{\mathrm{Tr}_{\mathrm{B}}\{e^{-\sum_{\alpha}\beta_{\alpha}(H_B^{\alpha}-\mu_{\alpha}N_{\alpha})}\}},
\end{equation}
where $\beta_{\alpha}=1/(k_{\rm B}T_{\alpha})$ denotes the inverse temperature with Boltzmann constant $k_{\rm B}$, $\mu_{\alpha}$ is the chemical potential and $N_{\alpha}=\sum_{k}c_{\alpha k}^{\dagger}c_{\alpha k}$ the  occupation number operator of lead $\alpha$, respectively. 
The difference of the chemical potentials of the left and right leads defines the bias voltage $\Phi$, which we assume to drop symmetrically, i.e., $\mu_{\rm L}=-\mu_{\rm R}=e\Phi/2$.

The corresponding extended state $|\Psi\rangle$ in twin space at the initial time is given by
\begin{equation}
    |\Psi(0)\rangle = |\underbrace{0 \cdots 0}_{K+2D}\rangle.
\end{equation}
Within the MPS/TT representation of $|\Psi(0)\rangle$, we have $A^{[i]}(0,0,0)=1$ and all other values are set to zero for each tensor $A^{[i]}$. It is noted that the assumption of a factorized form of the composite system density operator can be lifted by performing an imaginary time propagation beforehand, as proposed in our previous publication.\cite{Ke_J.Chem.Phys._2022_p034103}

For the results presented below, we assume that both left and right lead are initially in their thermal equilibrium at room temperature $T=300$ K, and the bias voltage $\Phi$ is applied symmetrically, i.e. $\mu_L = -\mu_R = \Phi/2$.  The molecule-lead coupling strength is fixed at $\Gamma=0.05$ eV. We adopt a large Coulomb interaction $U=8$ eV to fully suppress the population in the doubly occupied state within the bias voltage regime $\Phi=0-3$ V. The convergence is checked with respect to the number of Pad\'e poles, size of vibrational basis sets, time step, and maximal bond dimension, and the following values are used: $P=20$, $N_{\rm DVR}=80$ and $N_h=10$ (number of energetic eigenstates for the harmonic bending mode),  $\delta t=0.1$ fs, and the maximal bond dimension $r_{\rm max}=100$. 

For the diabatic coupling $\Delta(x_2)$ in \Eq{diabatic_term}, we find that introducing an exponentially decaying factor $e^{-\lambda x_2^2}$ with the damping parameter $\lambda$ in the above coupling format improves the convergence of the approach. A value of $\lambda = 1$ is used for all calculations presented below, based on tests to ensure that this additional decay factor does not influence the physical results (more details are provided in the supporting information (SI)).

\section{Results}\label{results}

In this section, we apply the methods introduced above to unravel the reaction mechanisms underlying the process of current-induced bond rupture in single-molecule junctions. To this end, we analyse the current-induced dissociation dynamics in a series of models with increasing complexity, as listed in Table \ref{model}. The first and second model consider only a single electronic state of the charged molecule and a single reaction coordinate. Thereby, electronic states of different character are considered, a $\pi^*$-state in Model \RNum{1} and  a $\sigma^*$-state with a purely repulsive PES in Model \RNum{2}. Such models have been investigated in detail in Refs. \onlinecite{Erpenbeck_Phys.Rev.B_2018_p235452,Erpenbeck_Phys.Rev.B_2020_p195421,Ke_J.Chem.Phys._2021_p234702}. In Model \RNum{3}, two electronic states of the charged molecule are considered, corresponding to a $\sigma^*$ and a $\pi^*$ state, respectively, however, without diabatic coupling between the states, i.e., $\Delta(x_2)=0$. Finally, Model \RNum{4} represents the complete model, described by the  Hamiltonian given in \Sec{Model}, including two electronic states of the charged molecule, two vibrational modes and a diabatic coupling between the two electronic states.


\begin{table}
\centering
\begin{tabular}	{ | p{3.5cm}  | p{0.7cm}| p{0.7cm}| p{0.7cm}| p{0.7cm} | }
 \hline
   Model                 & \hfil \RNum{1} &  \hfil  \RNum{2} & \hfil \RNum{3} & \hfil  \RNum{4} \\
 \hline
 $\pi^*$ state & \hfil \cmark  & \hfil  &  \hfil \cmark  & \hfil  \cmark \\

 $\sigma^*$ state &  \hfil  & \hfil \cmark  &  \hfil \cmark  &  \hfil \cmark \\
 diabatic coupling $\Delta(x_2)$      &   \hfil   & \hfil      &  \hfil    & \hfil  \cmark \\
 \hline
\end{tabular}
\caption{Four different models considered in the simulations. The first and second models consider only one electronic state of the charged molecule, of either $\pi^*$ or $\sigma^*$ character, respectively. The third model takes both charged electronic states into account. The fourth model includes, in addition, a diabatic coupling $\Delta(x_2)$ between two charged electronic states.}
\label{model}
\end{table}
\begin{figure}
\centering
\includegraphics[width=0.45\textwidth]{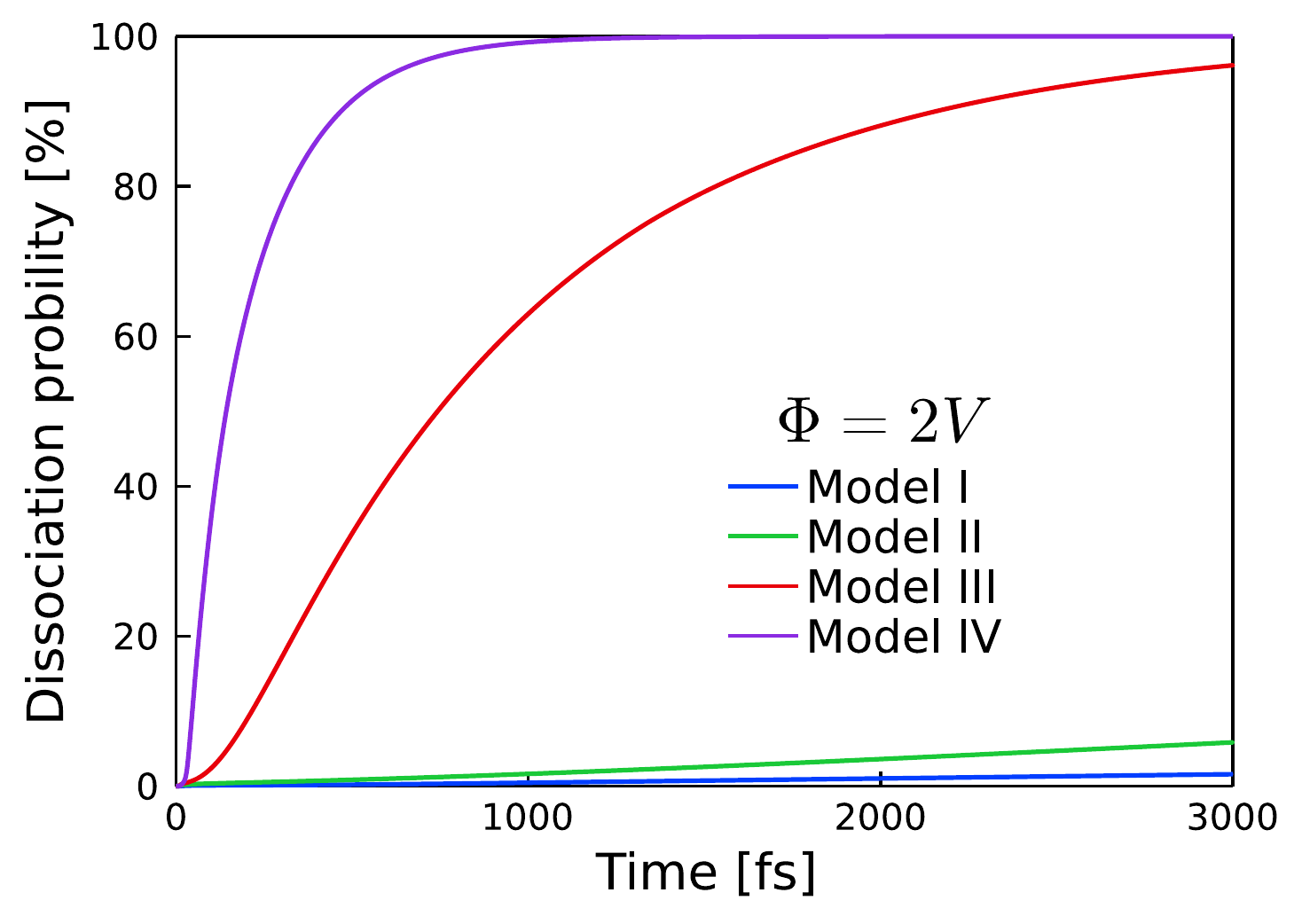}
\caption{Dissociation probability as a function of time at a bias voltage $\Phi=2$ V for four different models as listed in Table \ref{model}.}
\label{dissociation}	
\end{figure}

\subsection{Overview of dissociation mechanisms}\label{sec_mechanisms}

Experimental studies have found that most molecular junctions are unstable beyond a bias voltage of $1-2$ V.\cite{Sabater_BeilsteinJ.Nanotechnol._2015_p2338-2344,Li_J.Am.Chem.Soc._2016_p16159-16164}  In this section, we present and analyze the dissociation dynamics at a fixed bias voltage of $\Phi=2$ V. This representative parameter regime provides an overview of the different dissociation mechanisms. Results for other bias voltages,  $\Phi=1$ and $3$ V, are presented in the SI. 

\begin{figure*}
\centering
	  	\begin{minipage}[c]{0.23\textwidth} 		
			\raggedright a) Model \RNum{1}\\
	\includegraphics[width=\textwidth]{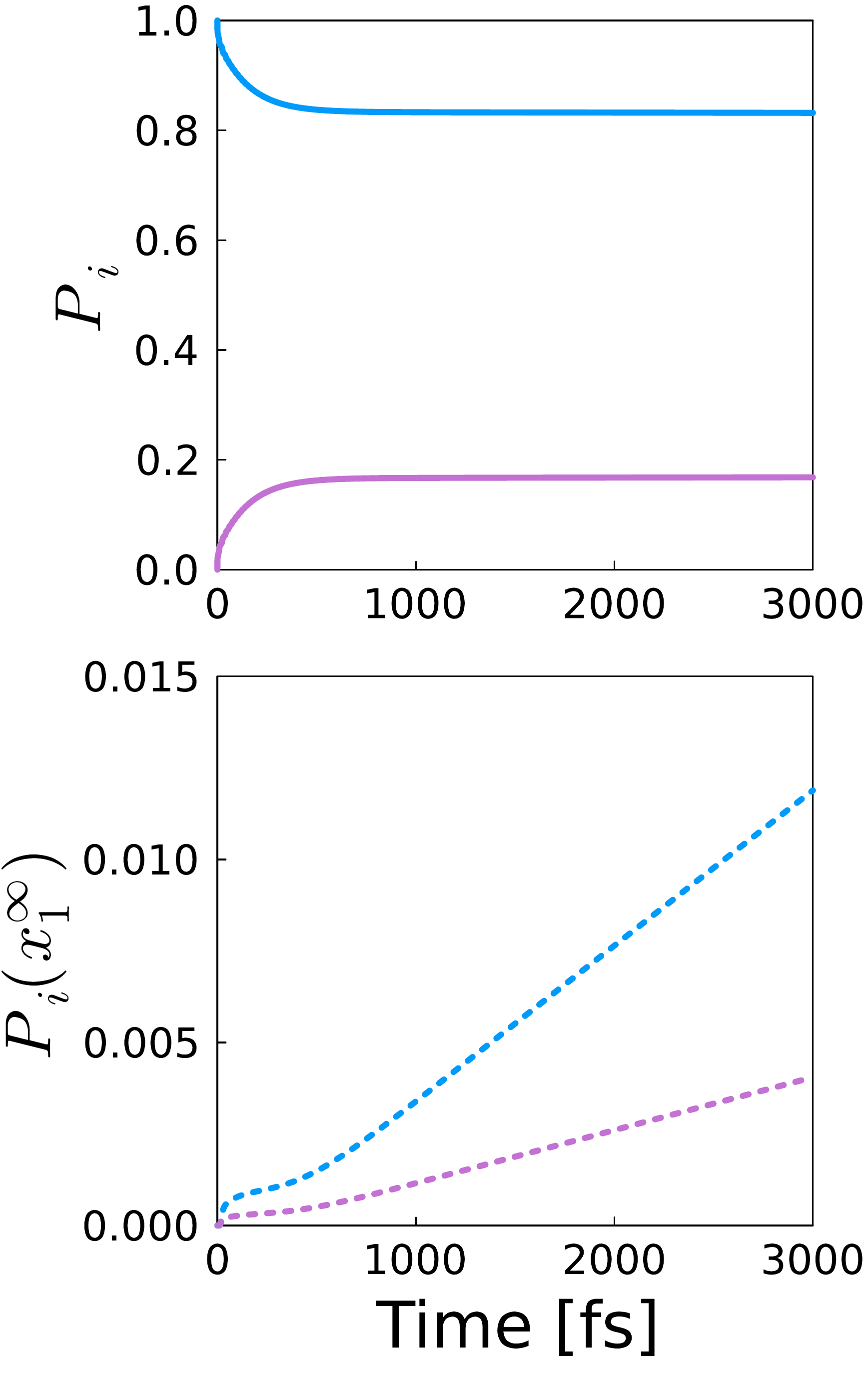}\
	\end{minipage}
	\begin{minipage}[c]{0.23\textwidth} 		
			\raggedright b) Model \RNum{2}\\
	\includegraphics[width=\textwidth]{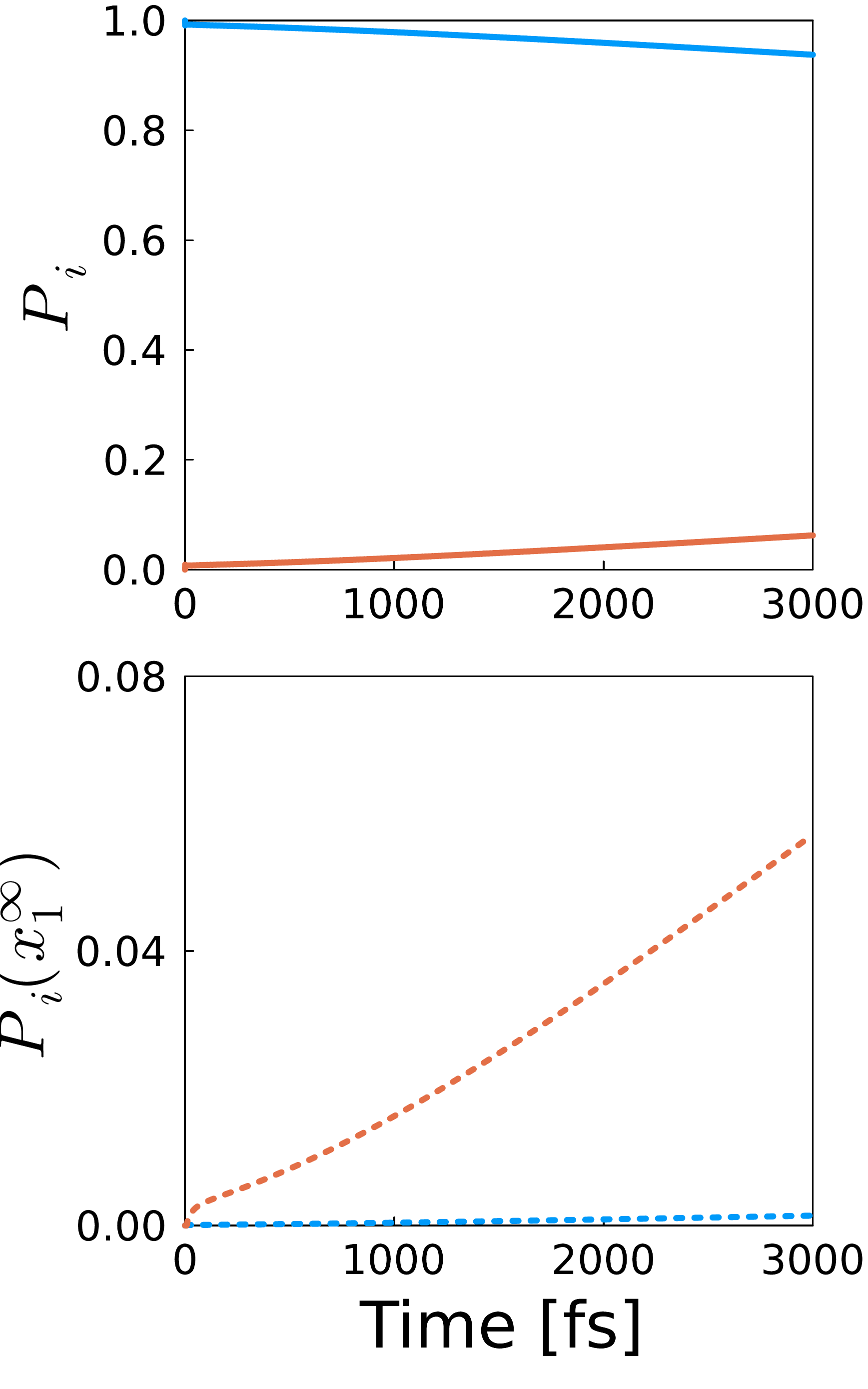}
	\end{minipage}
		\begin{minipage}[c]{0.23\textwidth} 		
			\raggedright c) Model \RNum{3}\\
	\includegraphics[width=\textwidth]{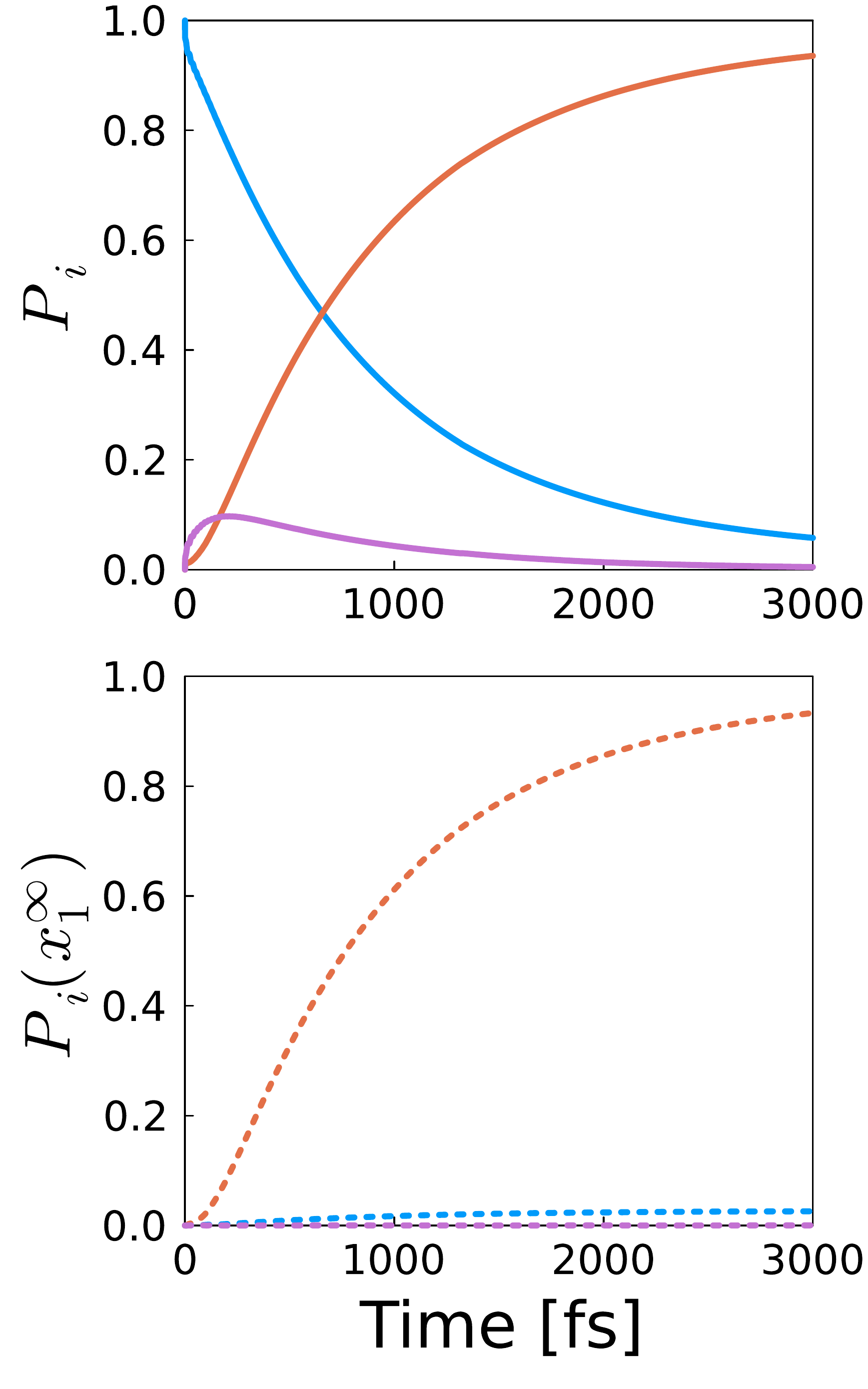}
	\end{minipage}
		\begin{minipage}[c]{0.23\textwidth} 		
					\raggedright d) Model \RNum{4}\\
	\includegraphics[width=\textwidth]{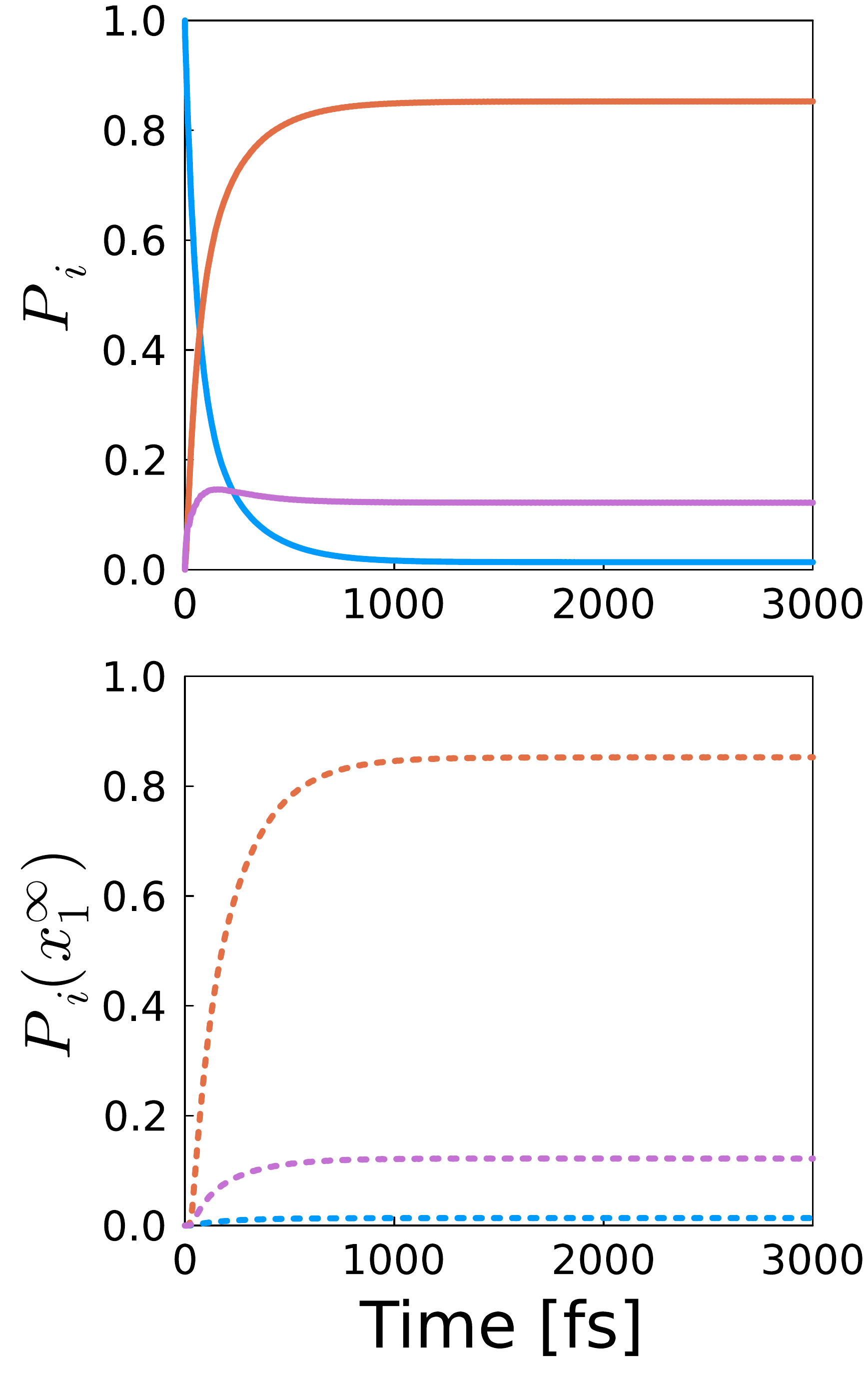}
	\end{minipage}		
 \begin{minipage}[c]{0.92\textwidth} 		
	\includegraphics[width=\textwidth]{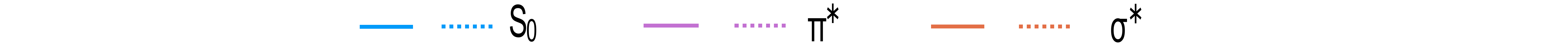}
	\end{minipage}
	  \caption{Population and dissociation dynamics in the different electronic states. Upper panels show the population of the electronic states ($P_i$ with $i\in \{S_0, \pi^*, \sigma^*\}$) as a function of time; Lower panels depict the electronic-state specific dissociation probability, given by the population at the point $x_1^{\infty}$ of the corresponding electronic states. The bias voltage is $\Phi=2$ V.  }
\label{elec_pop}	
\end{figure*}

\begin{figure*}
\centering
	\begin{minipage}[c]{0.23\textwidth} 		
		\raggedright a) Model \RNum{1}\\
	    \includegraphics[width=\textwidth]{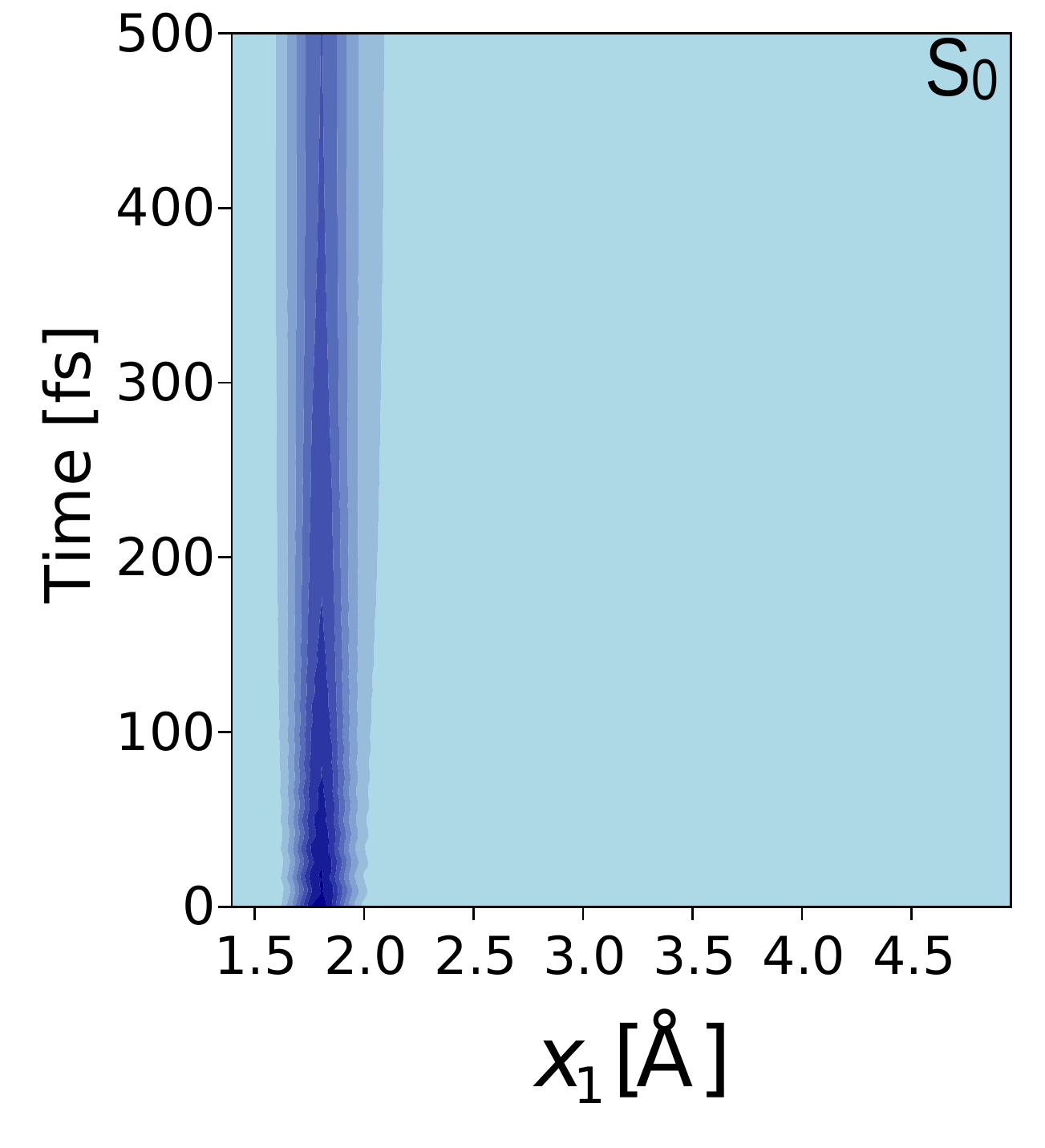}
	    \includegraphics[width=\textwidth]{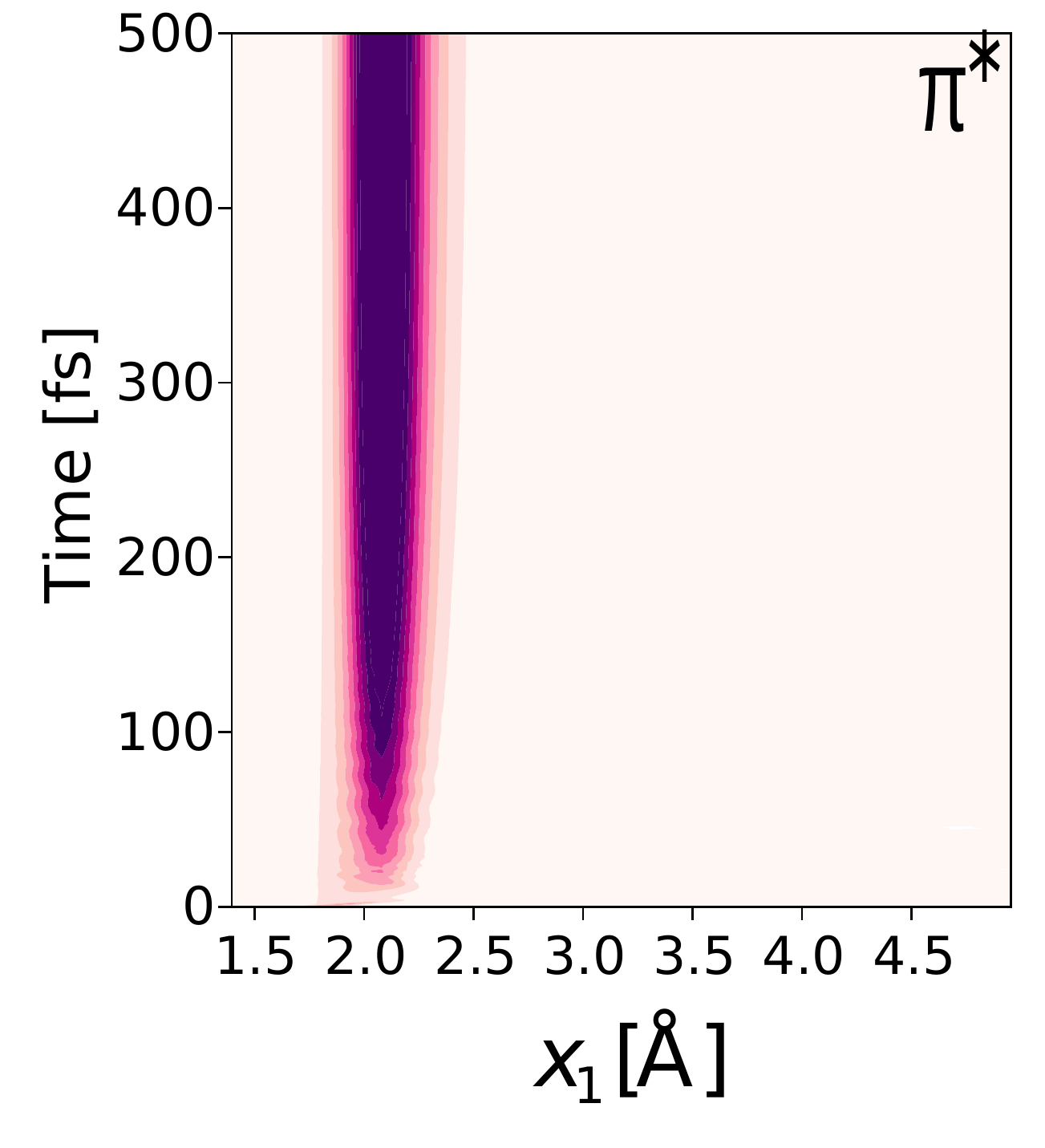}
	    \includegraphics[width=\textwidth]{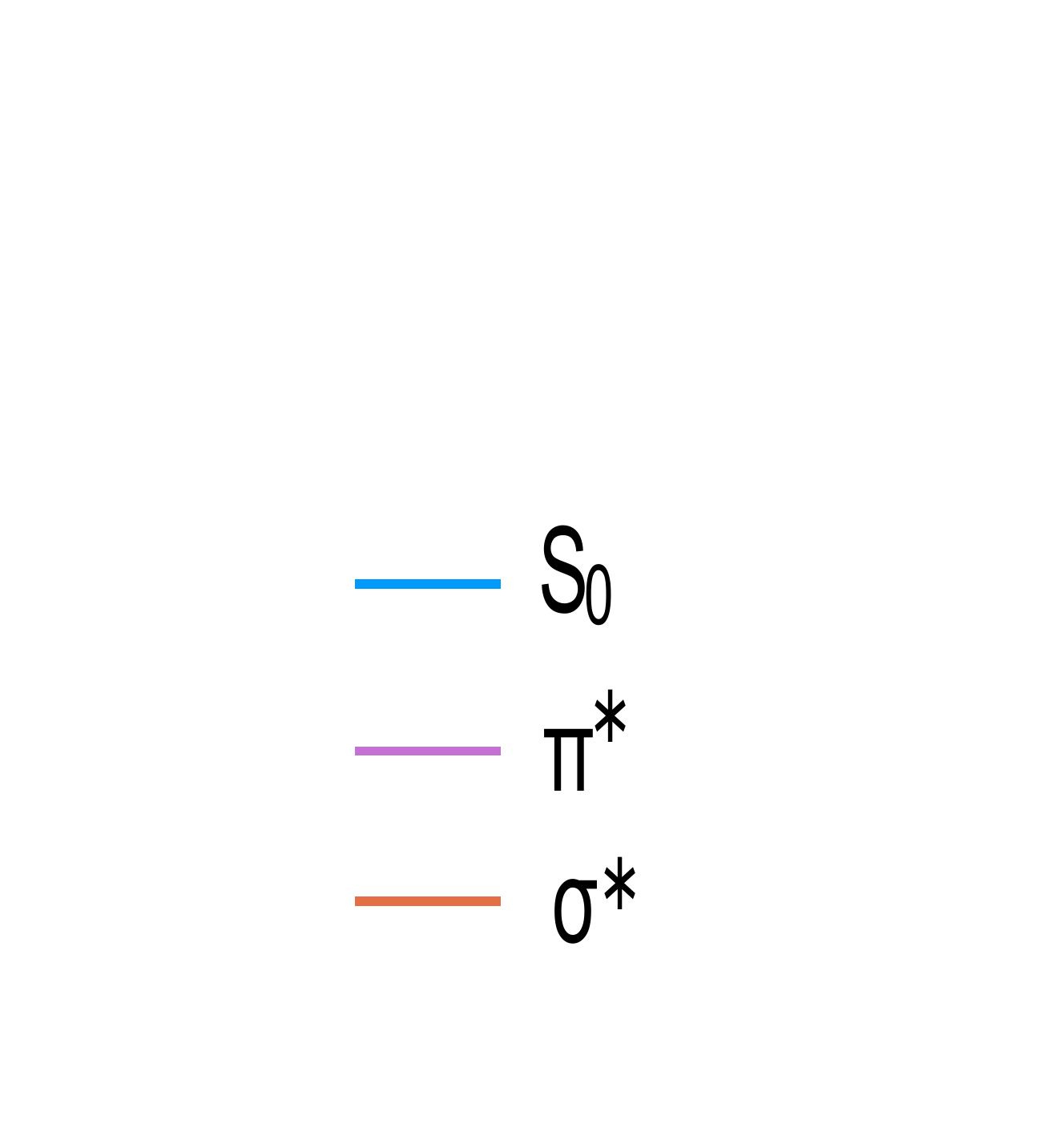}
		\includegraphics[width=\textwidth]{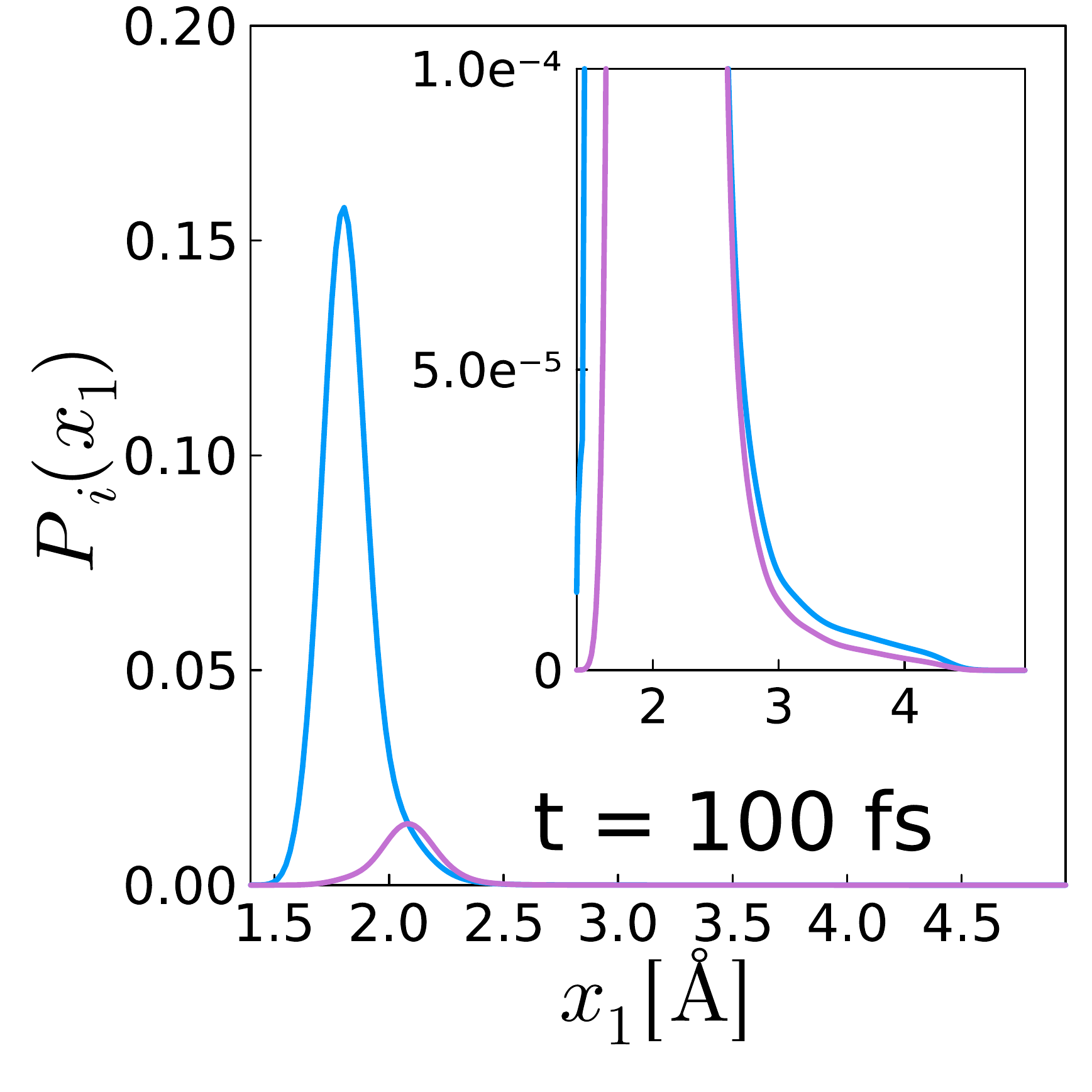}
	\end{minipage}
	\begin{minipage}[c]{0.23\textwidth} 		
		\raggedright b) Model \RNum{2}\\
	    \includegraphics[width=\textwidth]{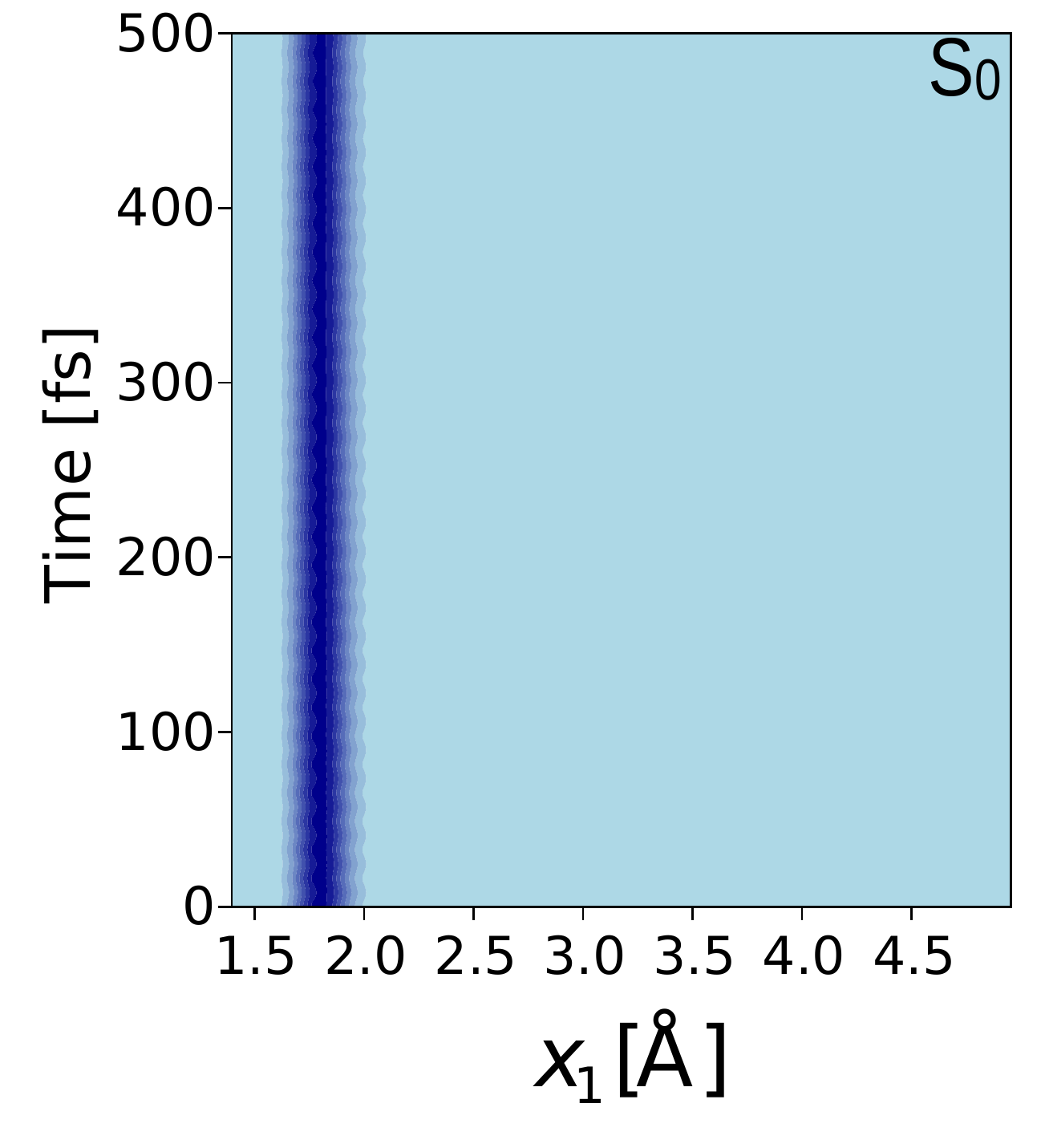}
	    \includegraphics[width=\textwidth]{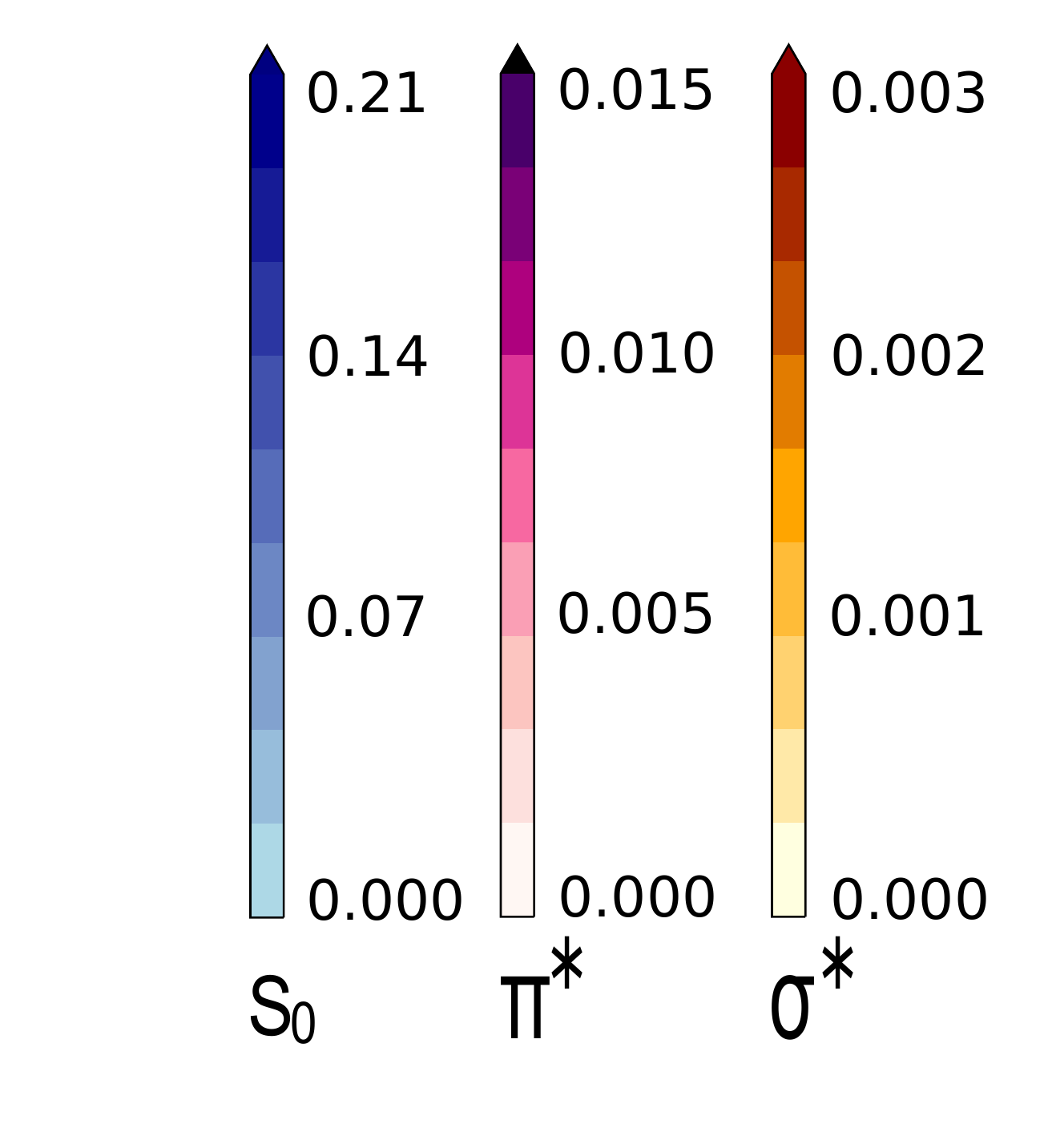}
	    \includegraphics[width=\textwidth]{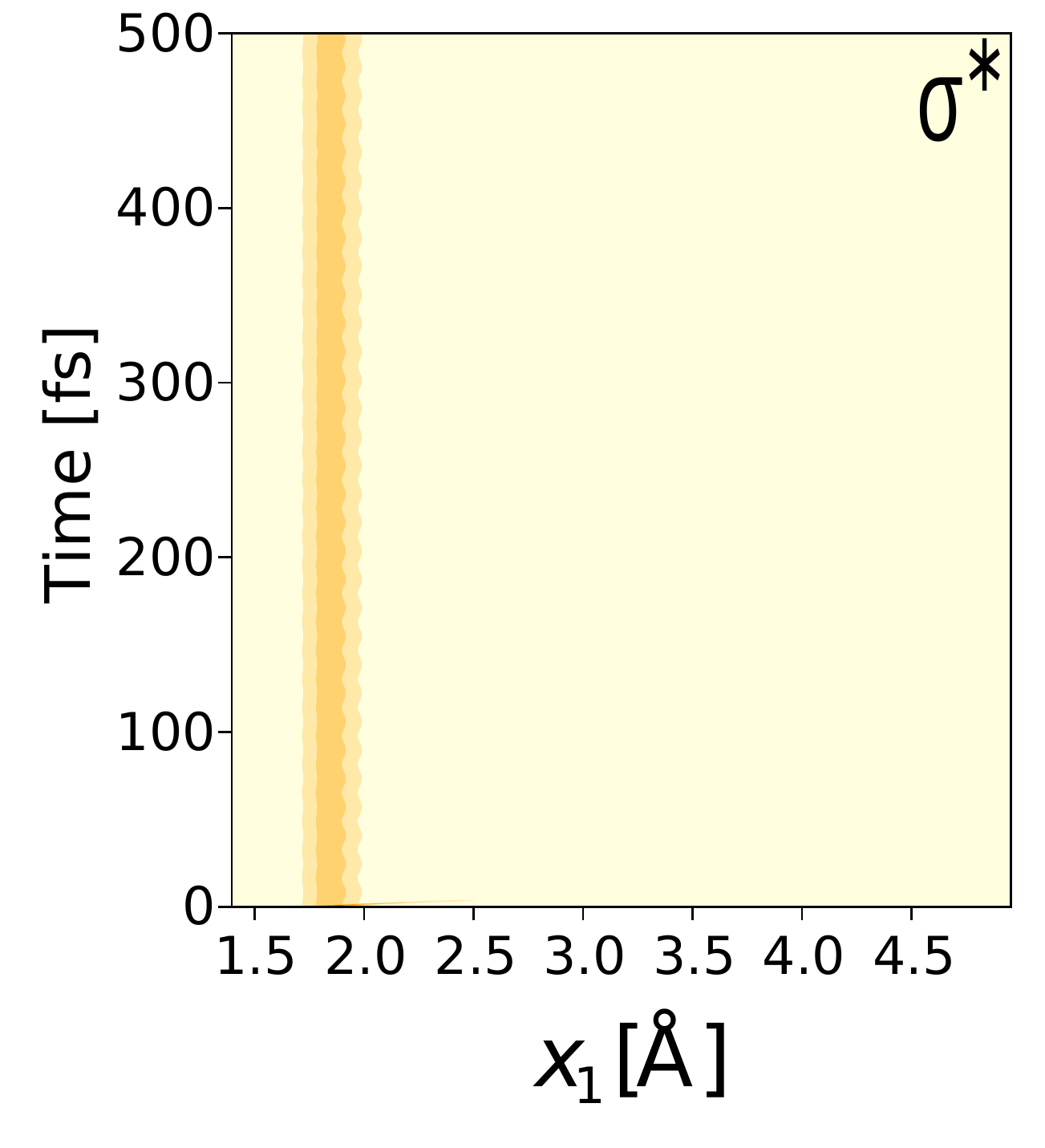}
	    \includegraphics[width=\textwidth]{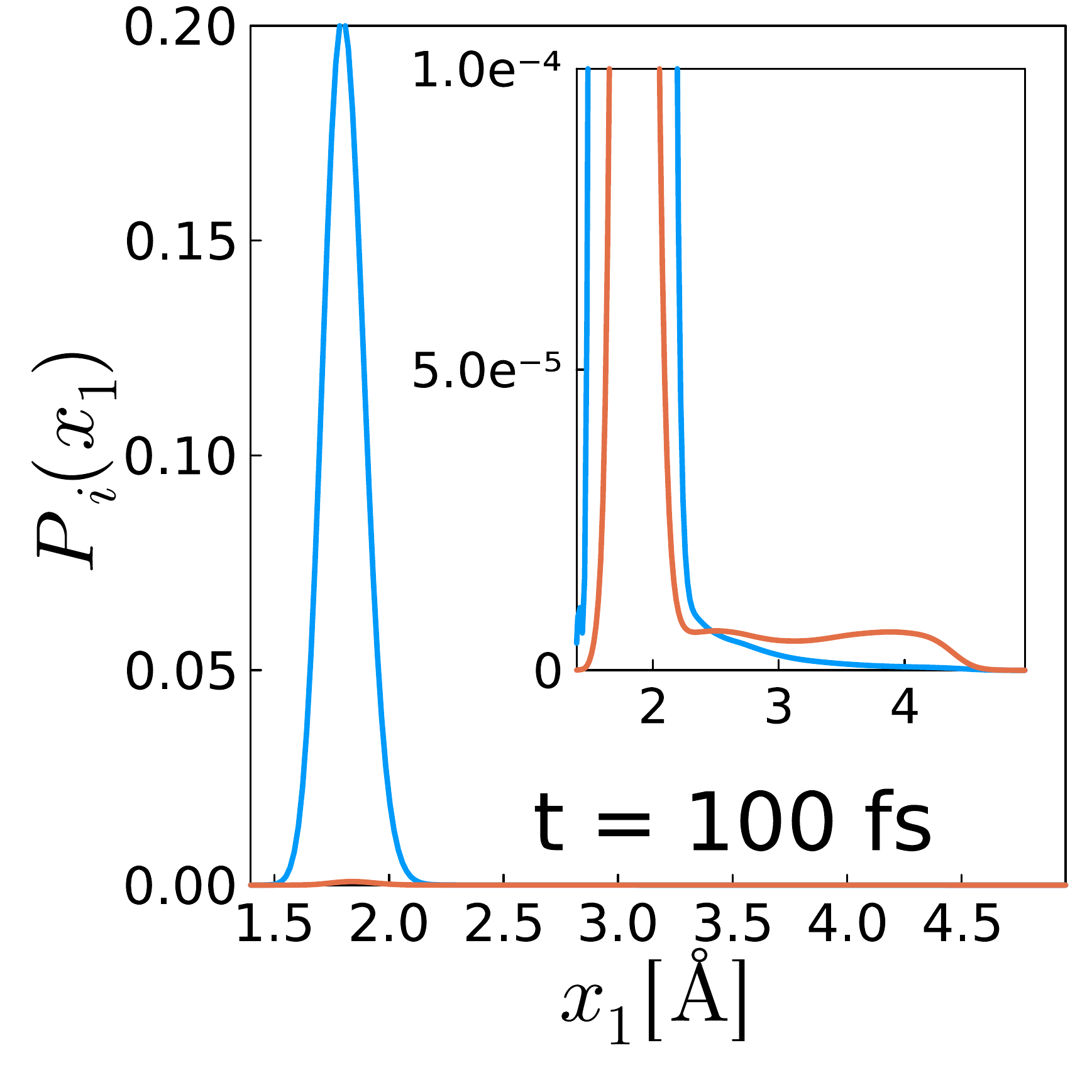}
	\end{minipage}
	\begin{minipage}[c]{0.23\textwidth} 		
		\raggedright c) Model \RNum{ 3}\\
	    \includegraphics[width=\textwidth]{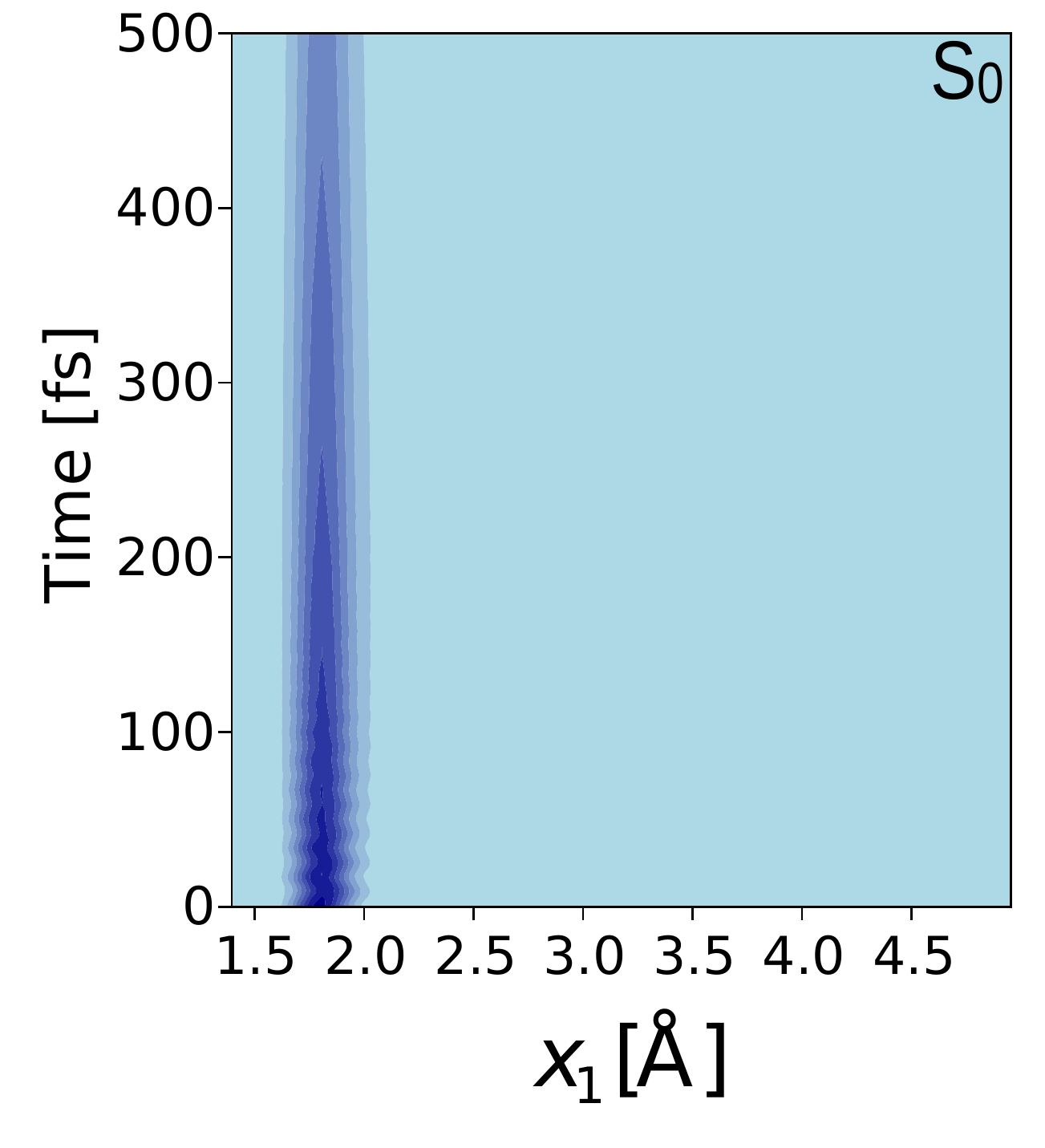}
	    \includegraphics[width=\textwidth]{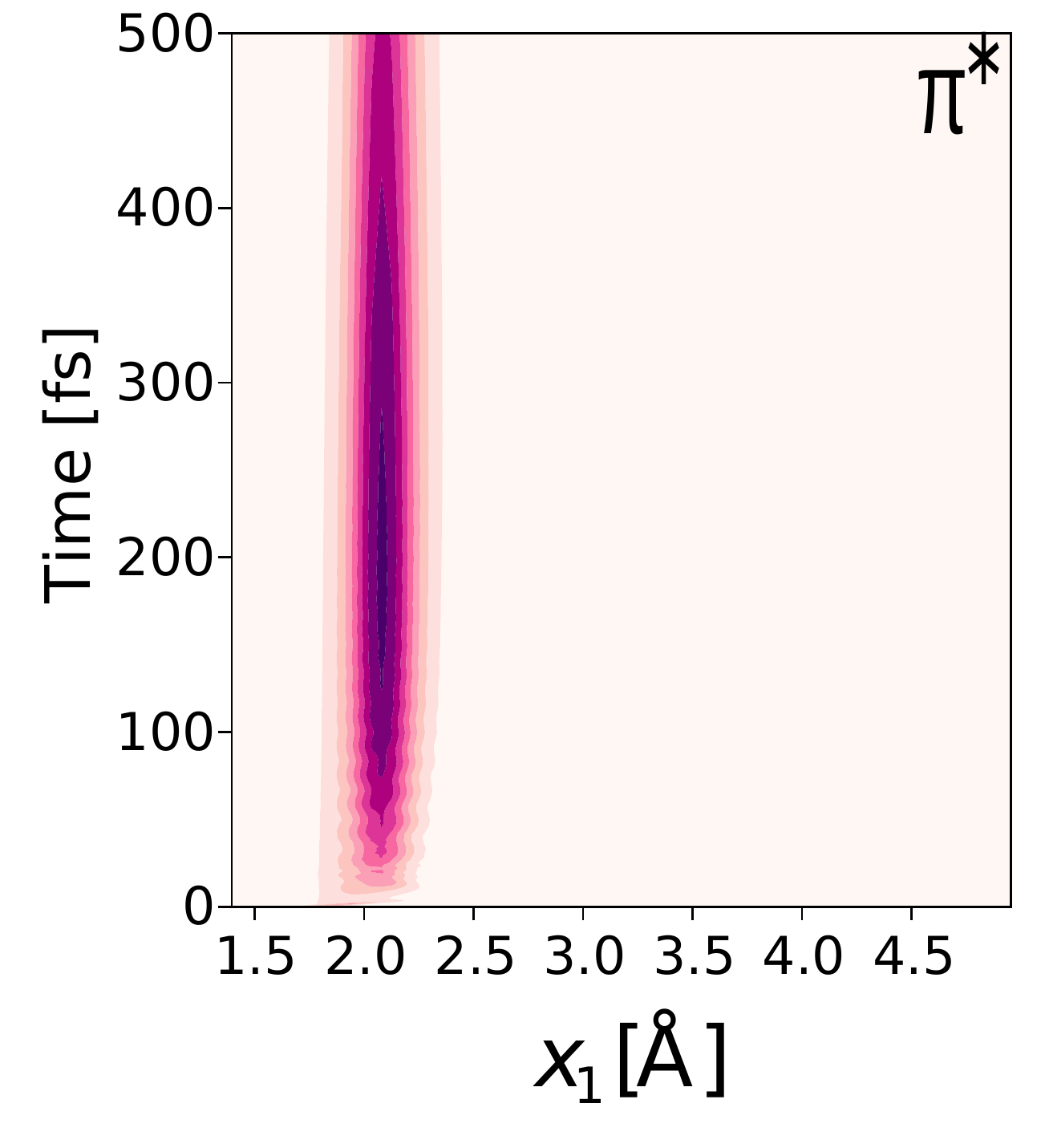}
	    \includegraphics[width=\textwidth]{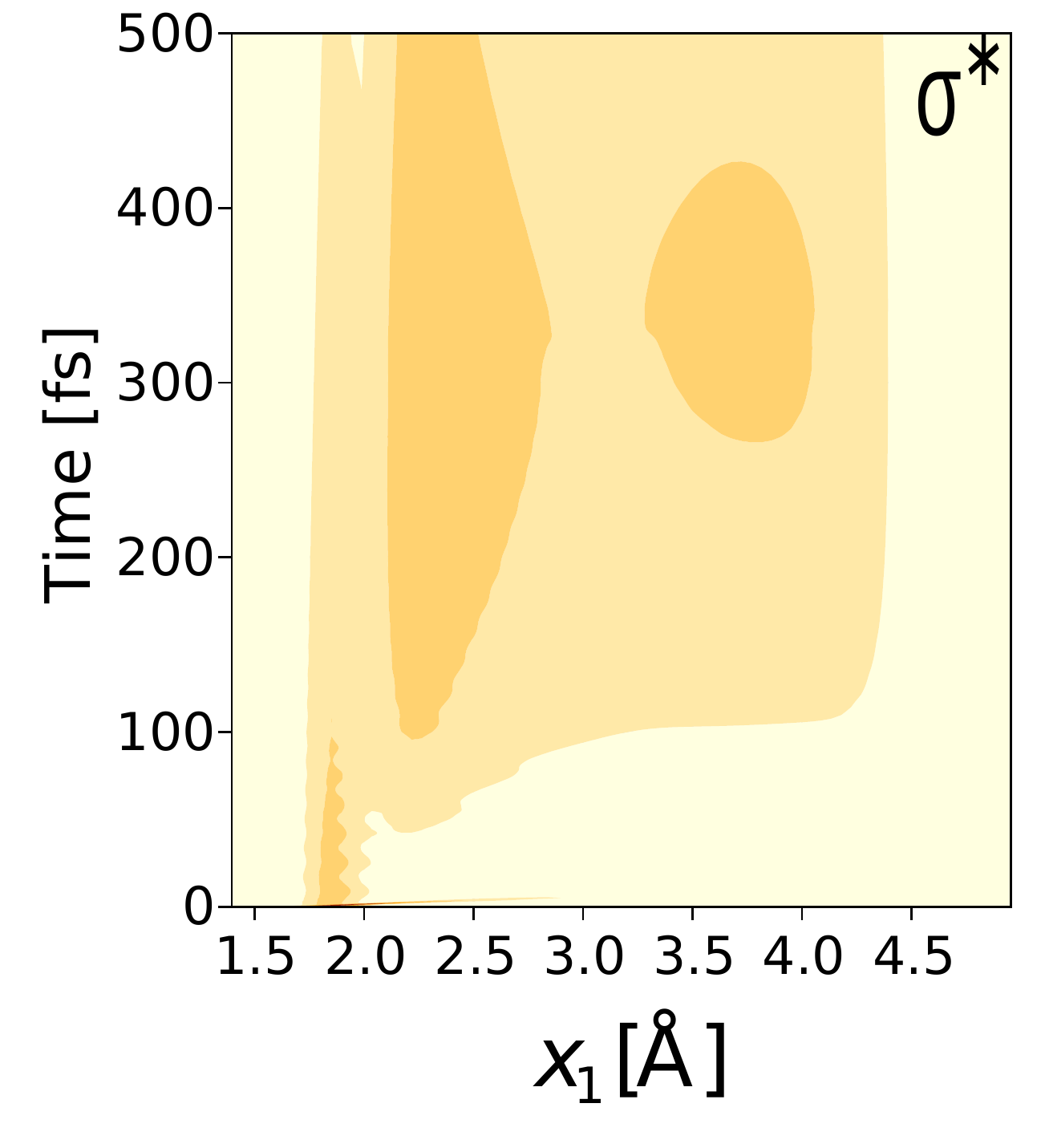}
	    \includegraphics[width=\textwidth]{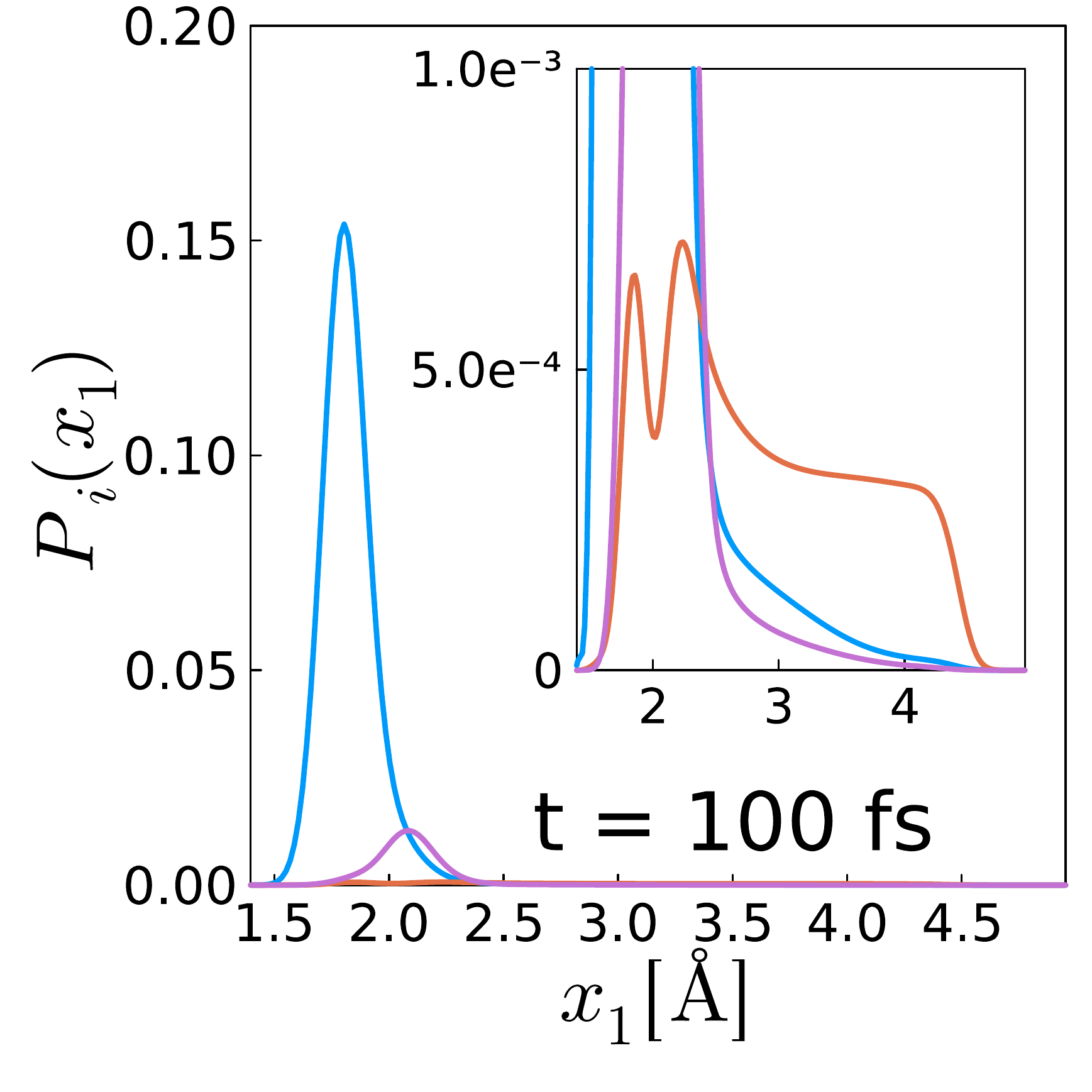}
	\end{minipage}
		\begin{minipage}[c]{0.23\textwidth} 		
		\raggedright d) Model \RNum{ 4}\\
	    \includegraphics[width=\textwidth]{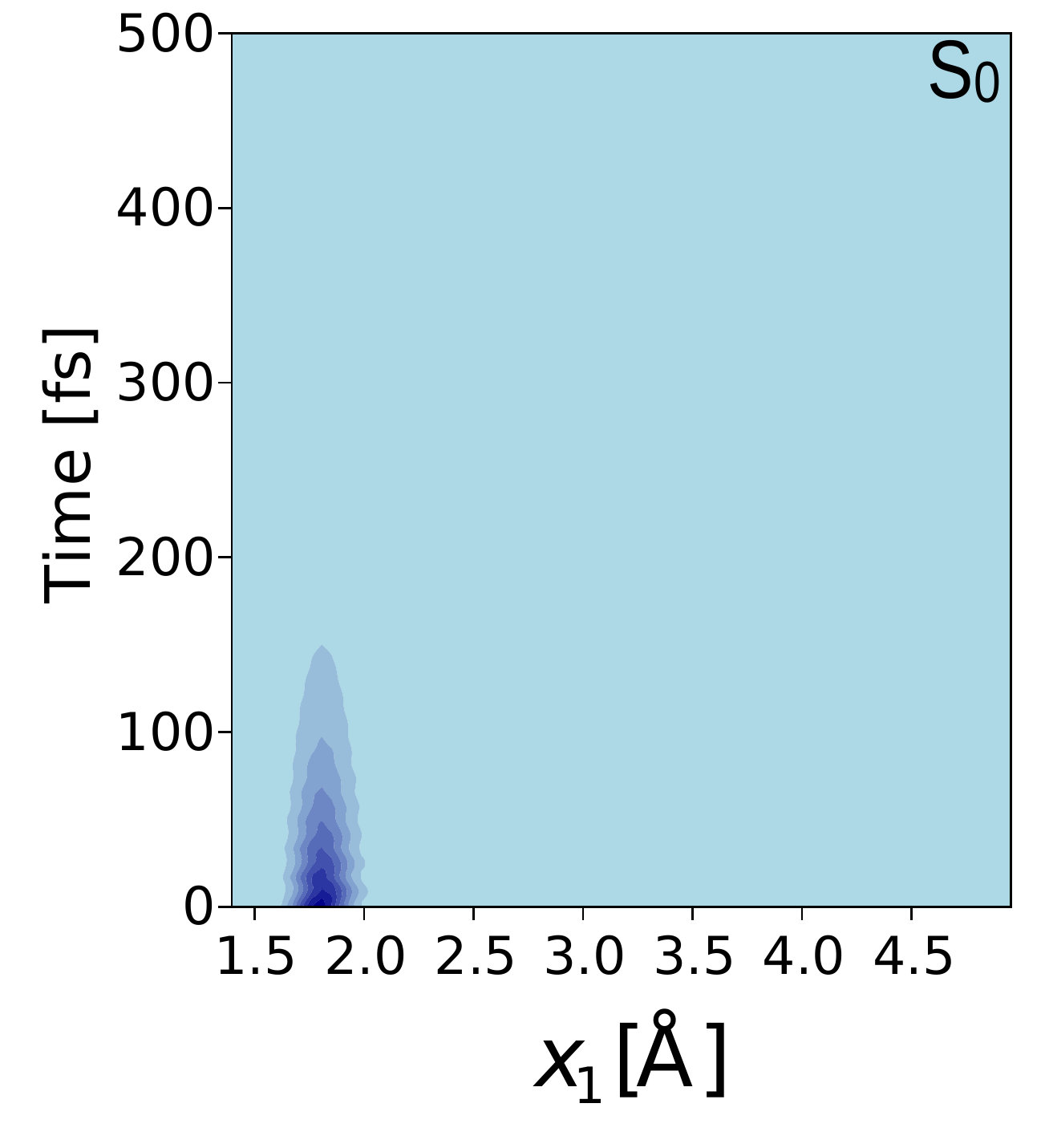}
	    \includegraphics[width=\textwidth]{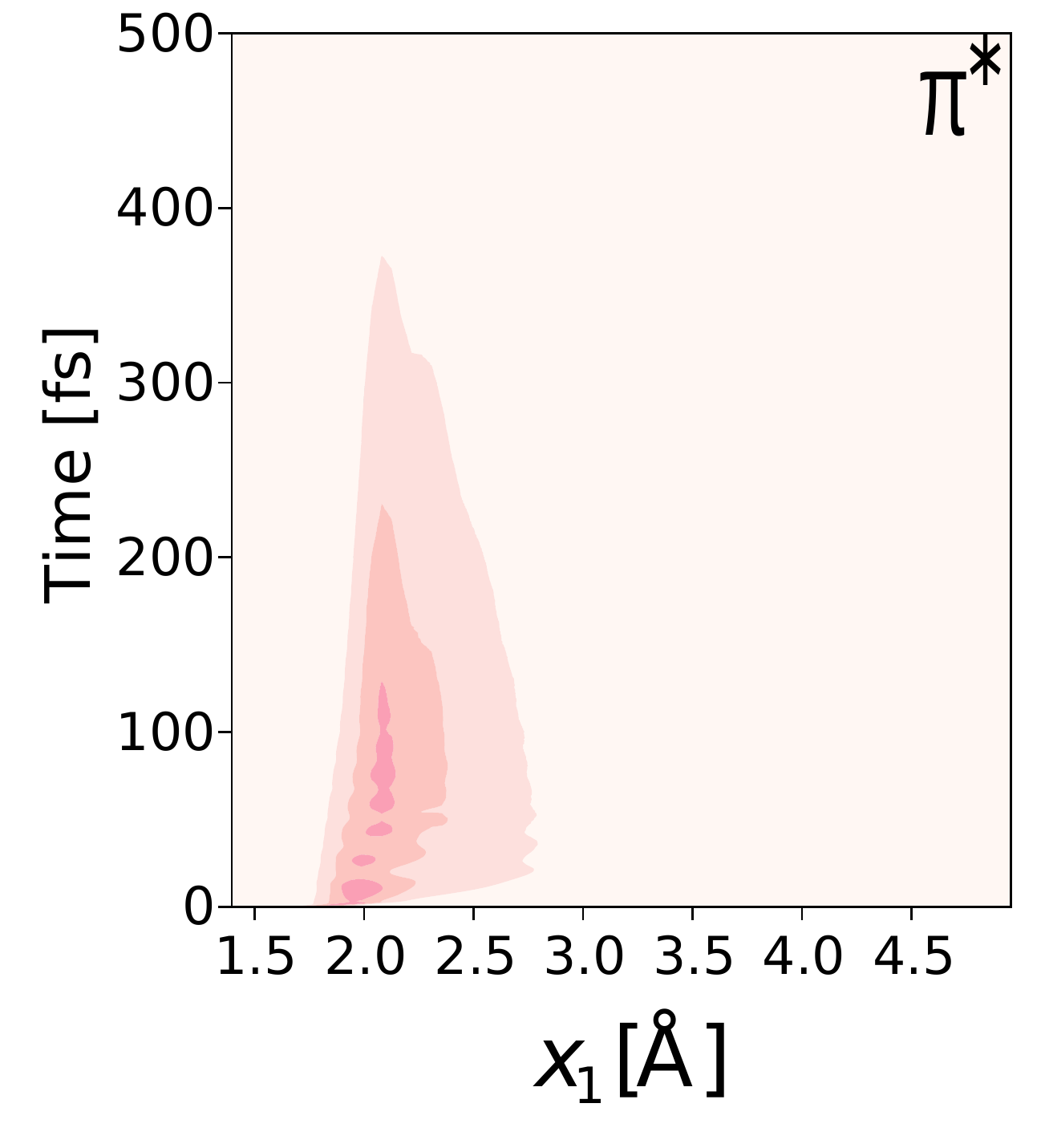}
	    \includegraphics[width=\textwidth]{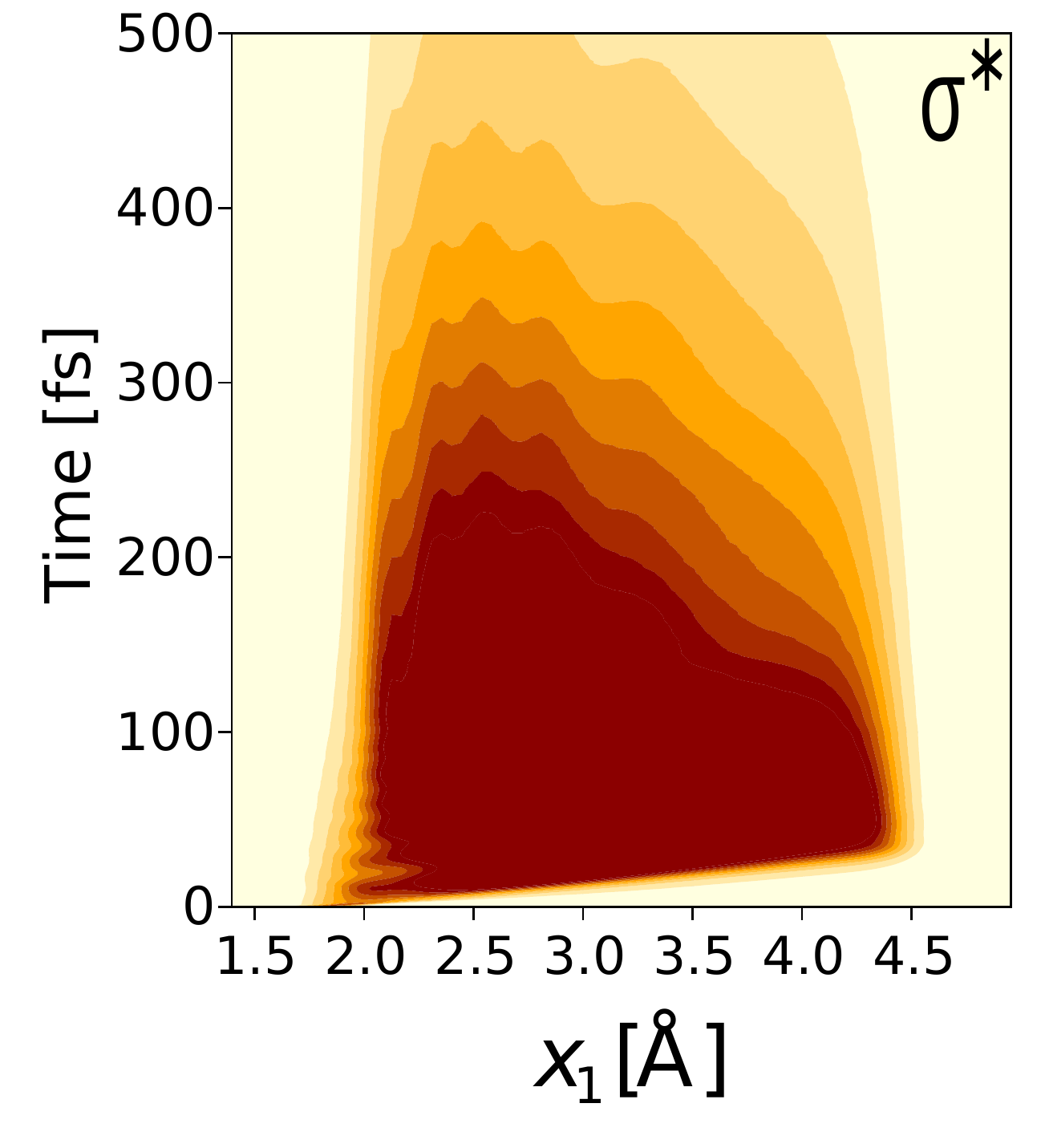}
	    \includegraphics[width=\textwidth]{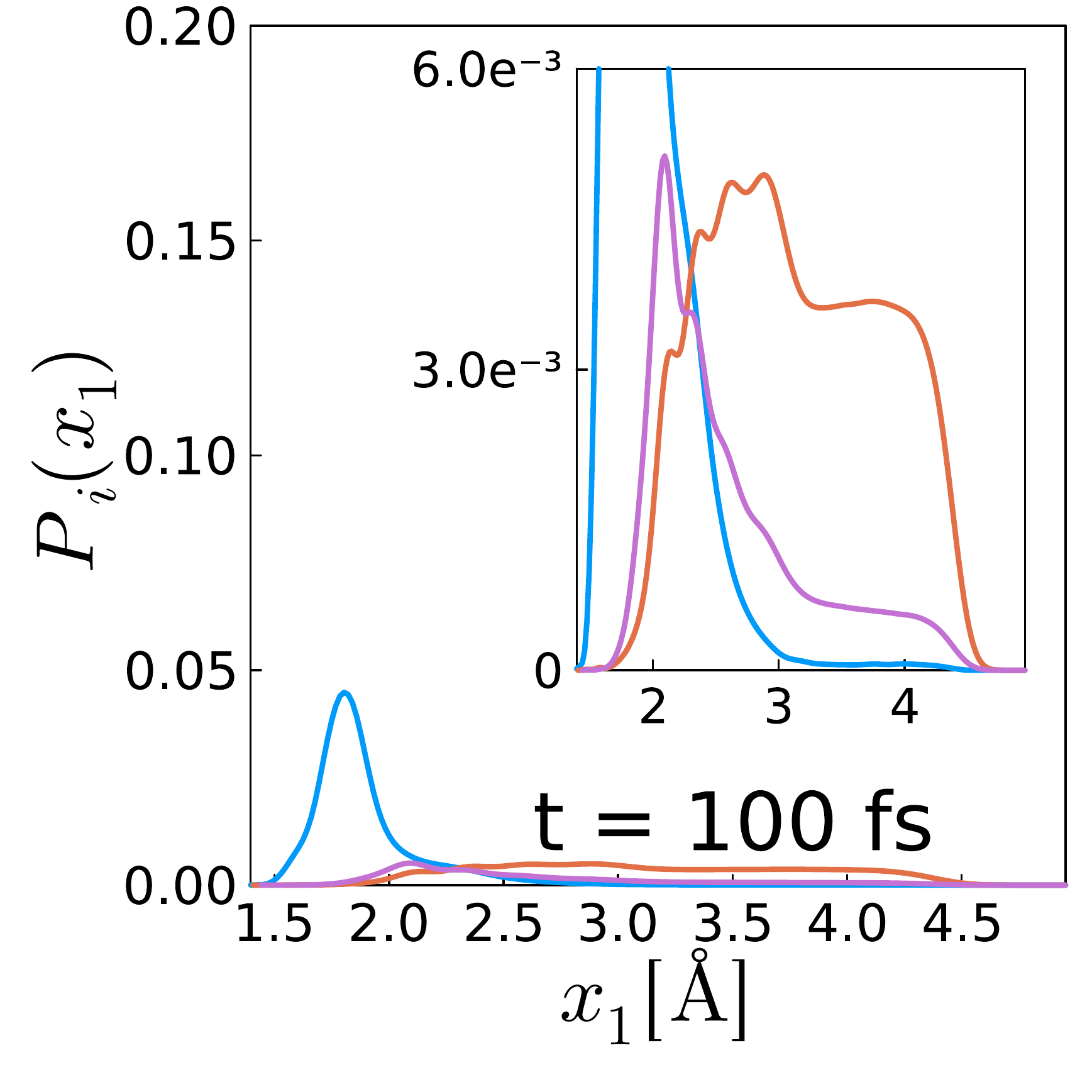}
	  \end{minipage}
	  \caption{Wave packet dynamics along the reaction mode $x_1$ in different electronic states (see \Eq{population_wavepacket}): neutral state $S_0$ (first row), $\pi^*$ charged state (second row), $\sigma^*$ charged state (third row).  Different columns correspond to different models as listed in Table \ref{model}. The panels in the fourth row show a snapshot of the wave packet dynamics at time $t=100$ fs. The blue lines are for the neutral state $S_0$ ($P_{S_0}$), magenta lines for the $\pi^*$ charged state ($P_{\pi^*}$) , orange lines for the $\sigma^*$ charged state ($P_{\sigma^*}$). To better resolve the dissociation dynamics, we also display a close-up of the wave packet in the small population regime in the inset of each panel. The bias voltage is $\Phi=2$ V. }
\label{pop_wavepacket_2V}	
\end{figure*}

\Fig{dissociation} displays the dissociation probability as a function of time for the four models. To facilitate the analysis of the underlying mechanisms, the electronic and vibrational dynamics are provided in \Fig{elec_pop} and \Fig{pop_wavepacket_2V}, respectively. 
In all simulations, the initial vibrational state is chosen as the vibrational ground state of the $S_0$ state of the neutral molecule.

\begin{figure*}
\centering
	\begin{minipage}[c]{0.24\textwidth} 		
			\raggedright a) Path \Circled{1}\\
	\includegraphics[width=\textwidth]{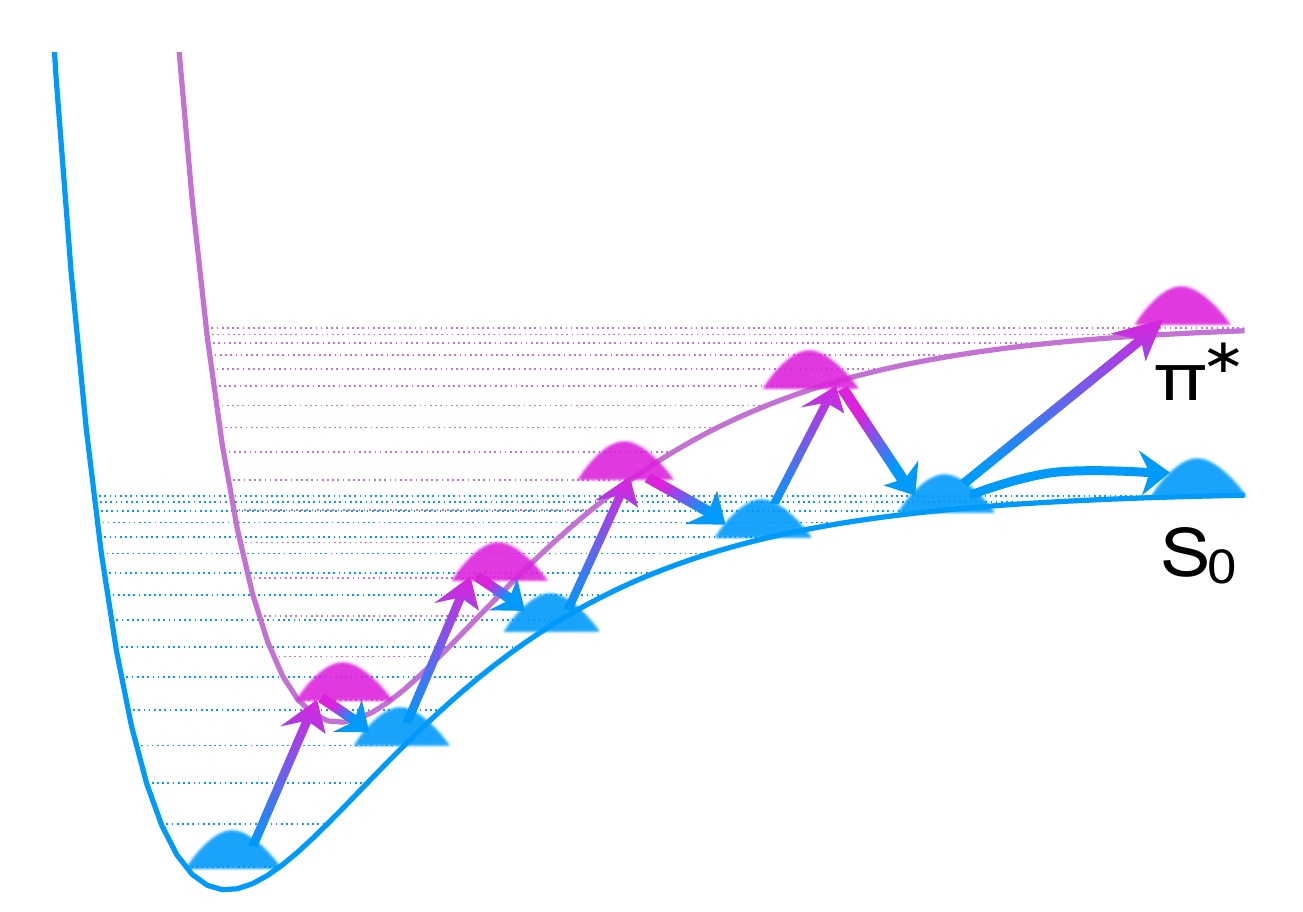}\\
	Current-induced vibrational ladder climbing
	\end{minipage}
	\begin{minipage}[c]{0.24\textwidth} 		
			\raggedright b) Path \Circled{2} \\
	\includegraphics[width=\textwidth]{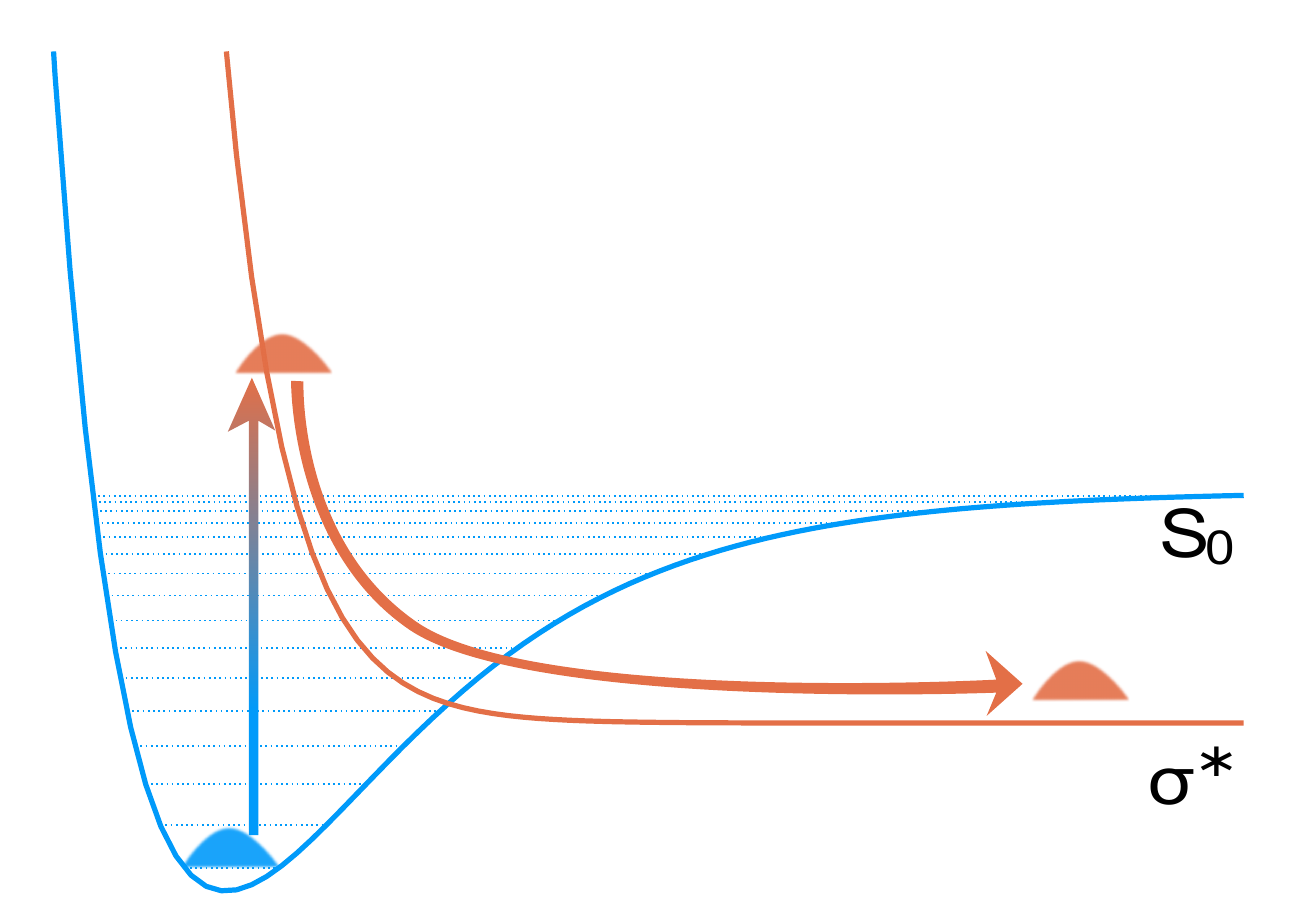}\\
	Direct dissociation in the repulsive state
	\end{minipage}
		\begin{minipage}[c]{0.24\textwidth} 		
			\raggedright c)  Path \Circled{3}\\
	\includegraphics[width=\textwidth]{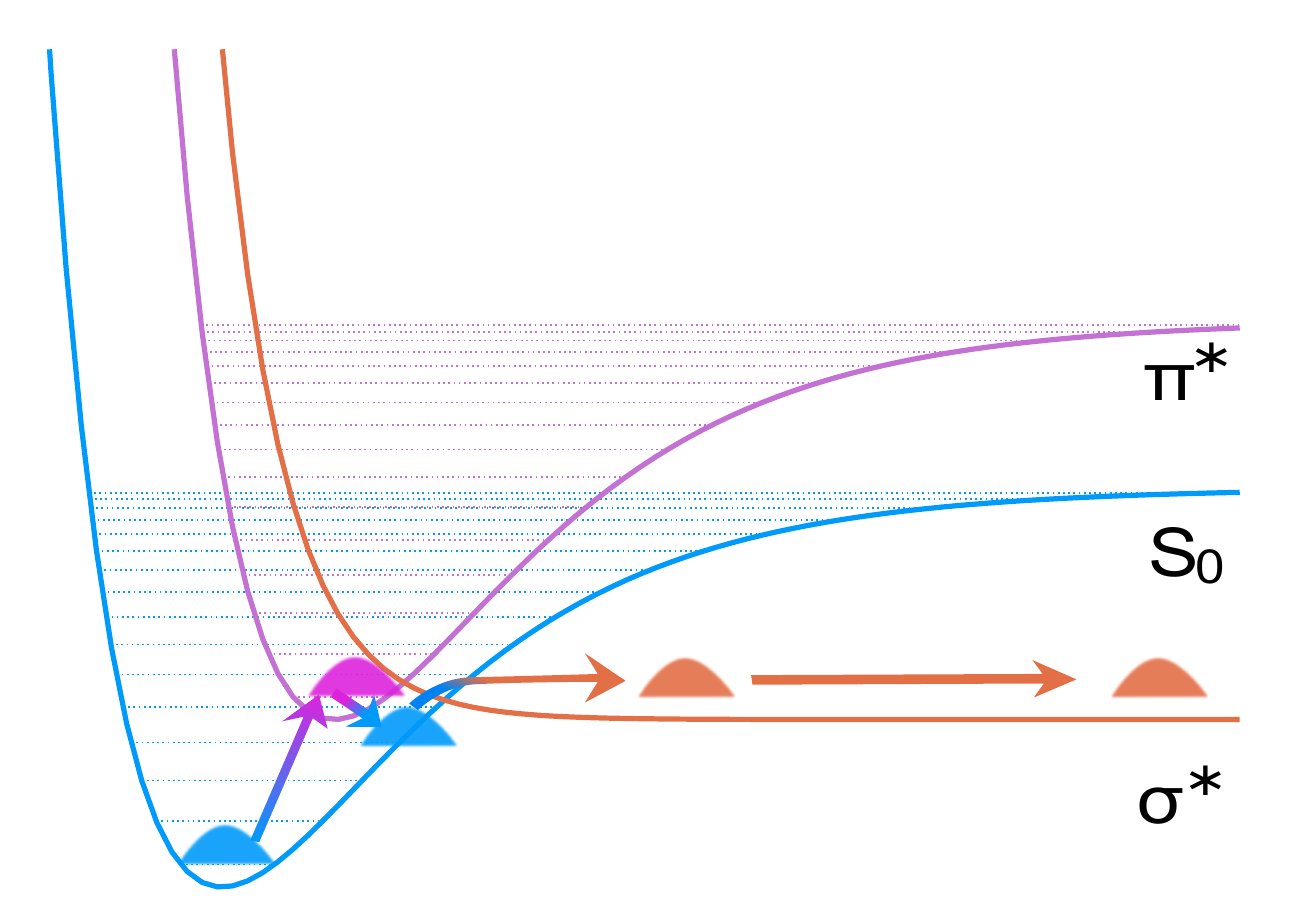}\\
	Heating-assisted direct dissociation
	\end{minipage}
		\begin{minipage}[c]{0.24\textwidth} 		
			\raggedright d) Path \Circled{4}\\
	\includegraphics[width=\textwidth]{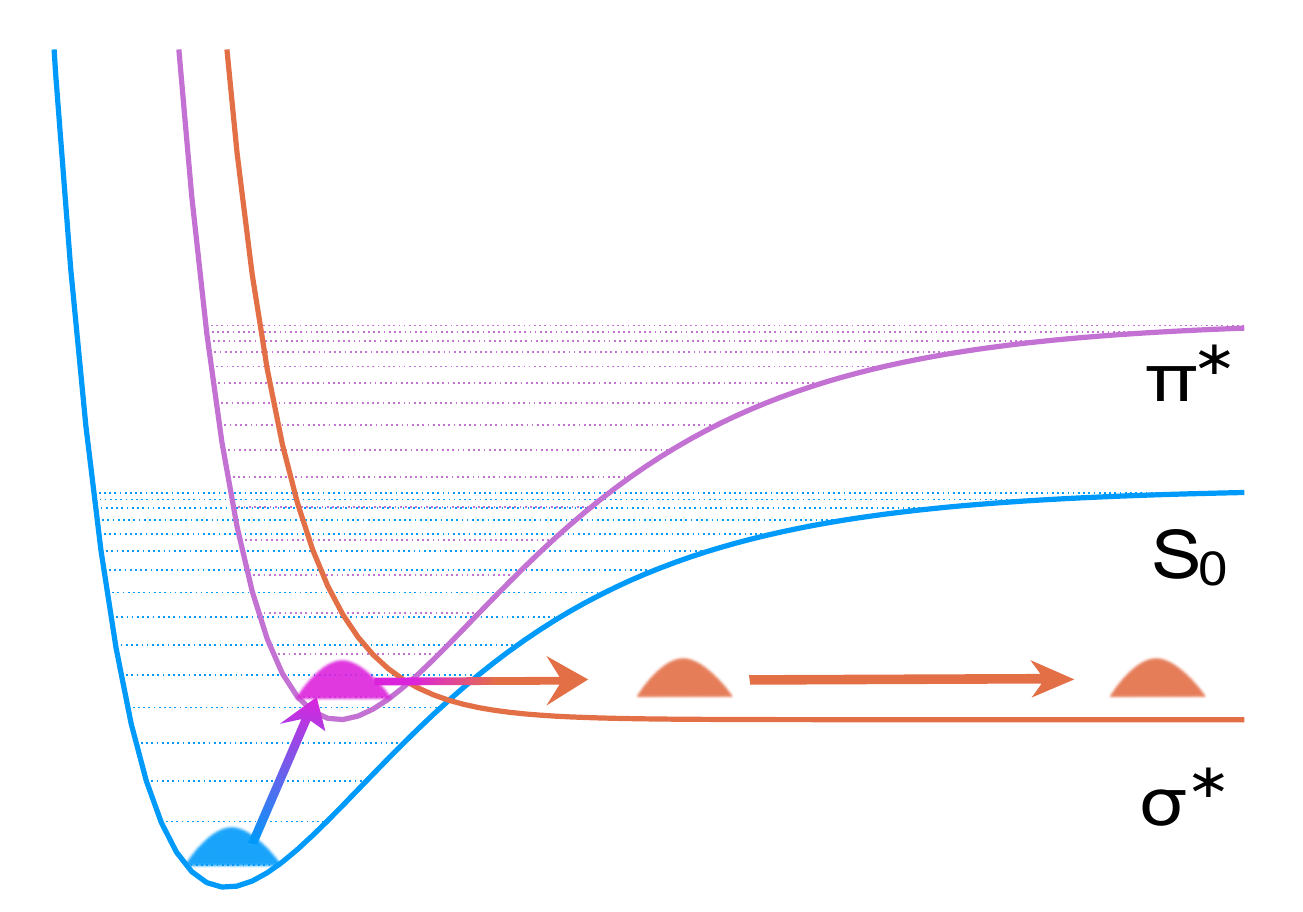}\\
	Dissociation induced by direct $\pi^*\rightarrow \sigma^*$ transition
	\end{minipage}
\caption{Schematic illustration of different dissociation mechanisms.  }
\label{mechanisms}	
\end{figure*}

Model \RNum{1} takes into account only the $\pi^*$ state of the charged molecule and the dissociative stretching mode $x_1$. The dissociation dynamics in \Fig{dissociation} shows for this model a non-zero but relatively small dissociation probability at long times. The electronic and vibrational dynamics in \Fig{elec_pop} (a) and \Fig{pop_wavepacket_2V} (a) reveal that a notable portion of the wave packet is transferred from the $S_0$ state of the neutral molecule  into the $\pi^*$ state of the charged molecule, and then quickly relaxes to the new equilibrium position centered at $x_1=2.08\text{ \AA}$, due to an efficient cooling effect caused by electron-hole pair creation processes. After a few hundred femtoseconds, the populations in the neutral and charged state reach a plateau. At the same time, caused by current-induced vibrational heating, the tail of the wave packet approaches the larger coordinate region, which eventually leads to dissociation. This dissociation pathway, corresponding to current-induced vibrational ladder climbing as schematically illustrated in \Fig{mechanisms} (a), requires multiple cycles of charging and discharging. Because the dissociation is induced by multiple electron attachment processes, the dissociation rate is relatively small, $k_{\text{diss}}=5.8*10^{-6}\text{ fs}^{-1}$.

In Model \RNum{2}, the considered charged state is of $\sigma^*$ character and, thus, purely repulsive. As a result, the dissociation takes place faster than in Model \RNum{1} with a rate of $k_{\text{diss}}=2.4*10^{-5}\text{ fs}^{-1}$ (see \Fig{dissociation}). The dissociation occurs exclusively in the charged state, as shown in \Fig{elec_pop} (b).  Inspecting the wave packet dynamics in \Fig{pop_wavepacket_2V} (b), we find that the wave packet remains largely in the neutral state and wiggles slightly forward and backward along the reaction coordinate with a period of 15 fs, which corresponds to the vibrational frequency estimated at the bottom of the Morse potential well. For the wave packet dynamics in the $\sigma^*$ charged state, we observe that, in addition to a small-amplitude main peak centered at the equilibrium position of the neutral state, there is also a broad and flat tail in the large reaction coordinate region.
In this case, the reaction is initiated by a vertical transition into the $\sigma^*$ state and followed by the outward diffusion in this repulsive surface, as demonstrated in  \Fig{mechanisms} (b).  

In Model \RNum{3}, where both charged states are involved but with a vanishing diabatic coupling, the dissociation is faster by several orders of magnitude ($k_{\text{diss}}=1.2*10^{-3}\text{ fs}^{-1}$), as shown in \Fig{dissociation}. This can be explained by the dissociation pathway, depicted in \Fig{mechanisms} (c). That is, the molecule is first heated to a low-lying vibrationally excited state of the neutral potential surface $S_0$ after a cycle of charging and discharging via the $\pi^*$ state. This excitation facilitates the transition from $S_0$ to the $\sigma^*$ charged state by the next incoming electron, because this process then only needs to overcome a very low or even no barrier.  Subsequently, the wave packet in the $\sigma^*$ state spreads quickly to the larger displacement region, driving a rapid dissociation. This heating-assisted direct dissociation process also explains the observations in the population dynamics, as shown in \Fig{elec_pop} (c). The population in the $\pi^*$ charged state is first increased in the short time regime and then drops to zero for longer times, as the $\pi^*$ charged state is an intermediate state for the preheating step. The wave packet dynamics of the $\sigma^*$ charged state (see \Fig{pop_wavepacket_2V} (c)) shows at short times a peak centered at $x_1^0=1.78\text{\AA}$, which as in Model \RNum{2} is caused by the vertical transition into the charged state. However, for times $t>50$ fs, a rising contribution at $x_1>2.0\text{\AA}$ is observed, peaking at the proximity of the $S_0$ and $\sigma^*$ crossing point.

Finally, we consider the most comprehensive model of our study, Model \RNum{4}, which comprises both charged states and the diabatic coupling $\Delta(x_2)$, which depends on the bending mode $x_2$. In this case, the dissociation is even faster ($k_{\text{diss}}=5.4*10^{-3}\text{ fs}^{-1}$) and completed within one picosecond, as shown in \Fig{dissociation}. This is because an additional dissociation channel is opened up, as depicted in \Fig{mechanisms} (d). The first step is the same as in Model \RNum{3}, starting from the vibrational ground state of the neutral state $S_0$, electron attachment promotes the wave packet into the lower-lying $\pi^*$ state. However, in the presence of a diabatic coupling, the wave packet can transfer directly from one surface to the other, i.e.\ undergo a $\pi^*\rightarrow \sigma^*$ transition.  As a consequence, a considerable population of the $\sigma^*$ state is already observed in a very short time, as shown in \Fig{pop_wavepacket_2V} (d). The wave packet in the $\sigma^*$ state then moves quickly along the repulsive surface towards the dissociation region. During this process, it is also possible that the wave packet transfers back to the $\pi^*$ state to the high-lying vibrationally excited states. The bond rupture occurs preferentially in the $\sigma^*$ state and only partially in the $\pi^*$ state, as shown in \Fig{elec_pop} (d).

Overall, the analysis reveals that there are four distinctively different dissociation mechanisms that can result in the bond cleavage, as illustrated in \Fig{mechanisms}. In the complete model accounting for all the relevant DoFs and their mutual interaction, we found that at a bias voltage of 2 V, the dissociation pathway through a direct $\pi^*\rightarrow \sigma^*$ transition is prevailing and dominates the reaction.  Nevertheless, as shown in our previous work,\cite{Ke_J.Chem.Phys._2021_p234702} the dominant reaction mechanism may depend sensitively on the applied bias voltage. Therefore, we proceed to study the dissociation dynamics over a range of bias voltage and determine the contributions of different mechanisms.

\subsection{Dissociation rate in different transport regimes}
\begin{figure}
\centering
	\begin{minipage}[c]{0.45\textwidth} 		
				\raggedright a) 
	\includegraphics[width=\textwidth]{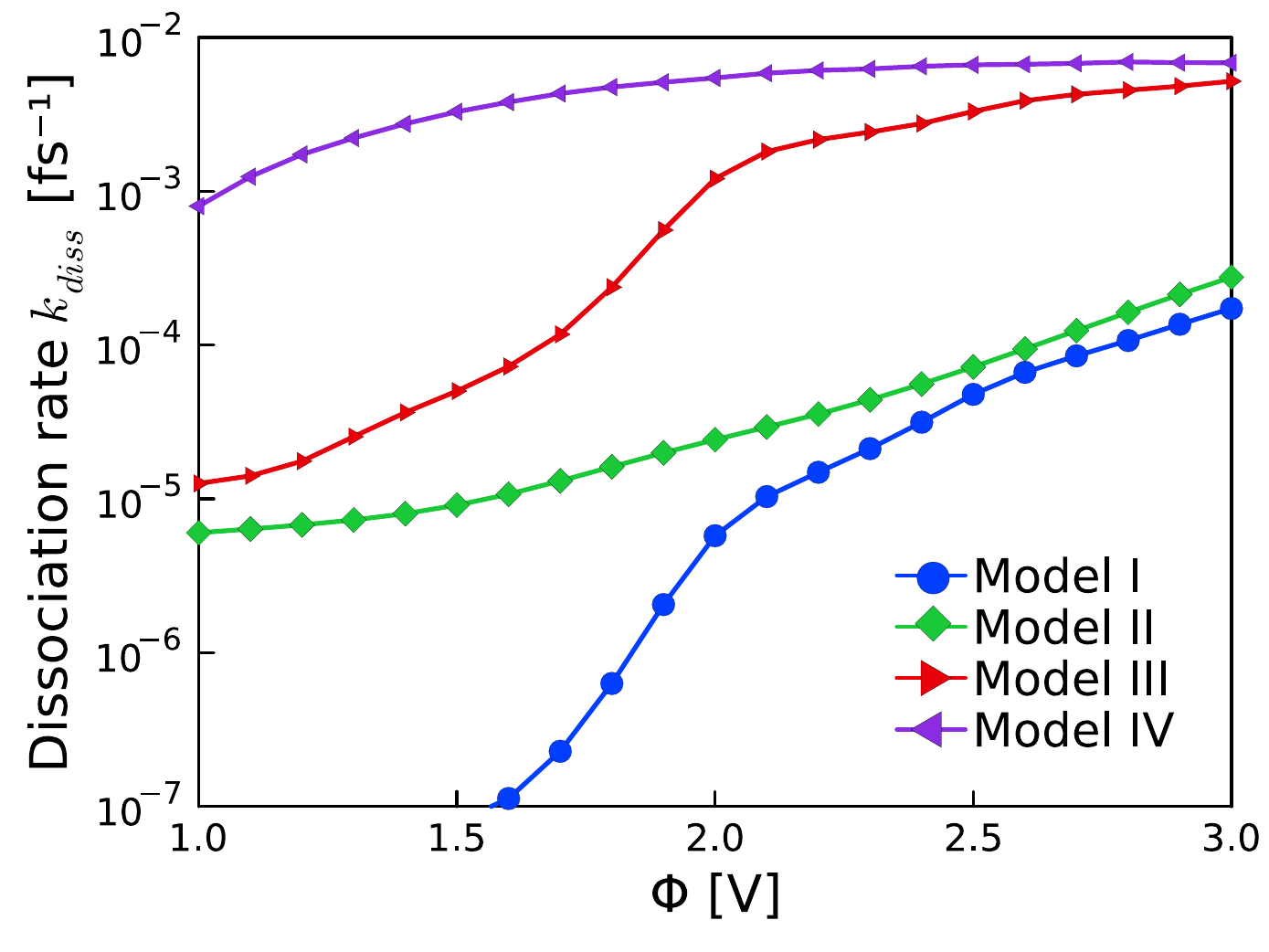}
	\end{minipage}
	\begin{minipage}[c]{0.45\textwidth} 		
		\raggedright b) 
	\includegraphics[width=\textwidth]{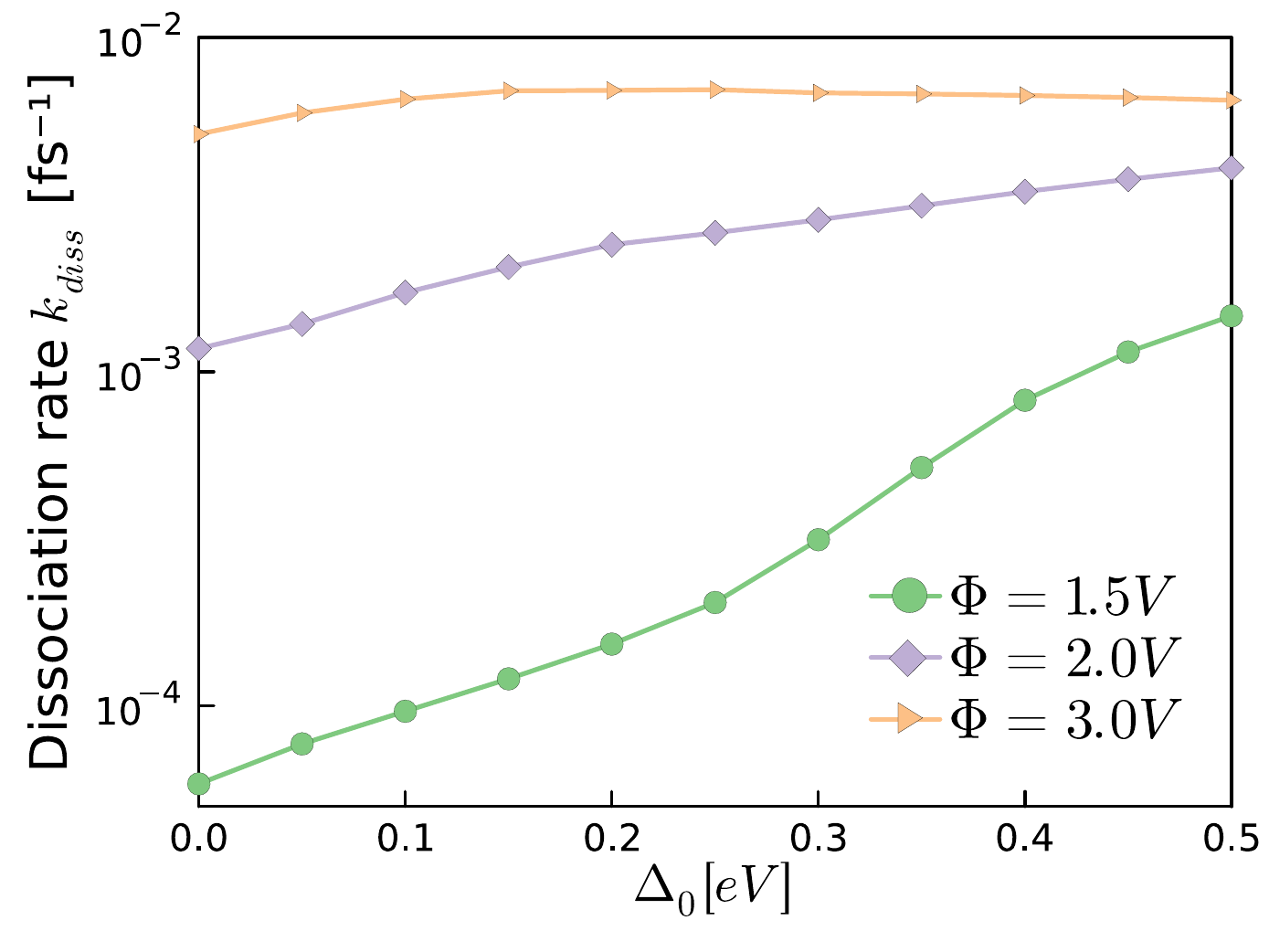}
	\end{minipage}
\caption{Dissociation rate $k_{\text{diss}}$ as a function of (a) the bias voltage $\Phi$ and (b) a constant diabatic coupling $\Delta_0$. In panel (a), different lines correspond to different models as listed in Table \ref{model}. In panel (b), different lines are for different bias voltages in a model with two charged states.}
\label{rates}	
\end{figure}

To gain more insight into the reaction mechanisms in different transport regimes, we display in \Fig{rates} (a) the dissociation rate $k_{\text{diss}}$ as a function of the bias voltage $\Phi$, ranging from 1 V to 3 V. 

For Model \RNum{1}, where current-induced vibrational ladder climbing (see Path \Circled{1} in \Fig{mechanisms} (a)) is the only possible dissociation mechanism, the curve of the dissociation rate $k_{\text{diss}}$ versus the bias voltage $\Phi$ exhibits two different slopes with a kink observed at around $\Phi=2$ V, which marks the transition from the non-resonant to the resonant transport regime. Below $\Phi=2$ V, the dissociation rate drops quickly with the decreasing bias voltage. At $\Phi<1.5$ V, the dissociation caused by current-induced heating is negligibly small on the simulated time scale (more details about the dissociation and population dynamics are provided in the SI).  Above $\Phi=2$ V, the $\pi^*$ charged state enters the resonant transport window set by the bias voltage, and the vibrational heating effect caused by inelastic transport processes becomes efficient. Increasing $\Phi$ to 3 V, the dissociation rate $k_{\text{diss}}$ is enhanced by an order of magnitude, and the dissociation occurs equally in the neutral $S_0$ and charged $\pi^*$ state.

In Model \RNum{2}, $k_{\text{diss}}$ is not negligibly small at low bias voltages ($\Phi<2$ V). For instance,   the rate at $\Phi=1$ V is $k_{\text{diss}}=6*10^{-6}\text{ fs}^{-1}$, only four times smaller than that at $\Phi=2$ V. Besides, we find that the dissociation takes place only in the $\sigma^*$ charged state, which indicates that the direct dissociation by charging into the repulsive state (see Path \Circled{2} in \Fig{mechanisms} (b)) is the dominant dissociation mechanism. At higher bias voltages, particularly when $\Phi>2.5$ V, the ratio of the dissociation in the neutral $S_0$ state is considerably increased (data not shown), suggesting that current-induced heating sets in.

Next, we turn to  Model \RNum{3}, which comprises two charged electronic states with vanishing diabatic coupling. As analyzed in \Sec{sec_mechanisms}, there is an extra dissociation channel, Path \Circled{3} in \Fig{mechanisms} (c), owing to a cooperative effect of vibrational heating and the ensuing transition to the dissociative $\sigma^*$ charged state. Therefore, the dissociation rate for Model \RNum{3} lies across the whole bias voltage range above that for Model \RNum{1} and Model \RNum{2}.  Even at $\Phi=1$ V, where the current-induced heating is inefficient, the dissociation rate is still doubled compared to that of Model \RNum{2}, which indicates that Path \Circled{3} contributes equally to the dissociation as Path \Circled{2}. This is because, in Path \Circled{3}, once the vibrational ladder climbing reaches 
the low-lying level at which the neutral state and the $\sigma^*$ charged state PESs intersect, the dissociation is then governed by a direct crossing to the dissociative $\sigma^*$ charged state with little or no activation energy. At higher bias voltages ($\Phi>2$ V) where current-induced heating becomes efficient,  Path \Circled{3} quickly dominates over the other two dissociation mechanisms. As such, the dissociation rates for Model \RNum{3} are over an order of magnitude larger than that for Models \RNum{1} and \RNum{2}. 

In the presence of a diabatic coupling $\Delta(x_2)$, which is the case for Model \RNum{4}, the dissociation rate is generally larger than that of the other models. At $\Phi=1$ V, the dissociation rate is already relatively high, $k_{\text{diss}}=8*10^{-4}\text{ fs}^{-1}$. With increase of the bias voltage, the dissociation rate is first increased, but then gradually levels out. To analyze the impact of the diabatic coupling per se, \Fig{rates} (b) shows the dissociation rate as a function of a constant diabatic coupling $\Delta_0$ for different bias voltages. We note in passing that, the respective influence of the constant and coordinate-dependent terms in \Eq{diabatic_term} on the dissociation dynamics are discussed in the SI. At low bias voltages in the non-resonant transport regime, the rate increases pronouncedly for a larger $\Delta_0$, as the dissociation is predominantly driven by a direct $\pi^*\rightarrow \sigma^*$ transition  (see Path \Circled{4} in \Fig{mechanisms} (d)), whose timescale is inversely proportional to the diabatic coupling strength. 
It is also confirmed in \Fig{rates} (b) that, at high bias voltages in the resonant transport regime, the reaction rate is only weakly affected by a large diabatic coupling. This is because, at higher bias voltages, the heating-assisted direct dissociation (Path \Circled{3} in \Fig{mechanisms} (c)) becomes more important and fast enough to be comparable with the dissociation path induced by the direct $\pi^*\rightarrow \sigma^*$ transition, which then attenuates the role played by the diabatic coupling. 

Overall, the above analysis reveals that the consideration of multiple electronic states and vibrational modes is crucial to understand current-induced bond rupture mechanisms in molecule junctions. 

\section{Conclusions}\label{conclusion}
We have investigated current-induced bond rupture in molecular junctions employing the HEOM+MPS/TT method, which allows an accurate, fully quantum mechanical simulation of this challenging nonequilium quantum transport problem. Extending previous work,\cite{Erpenbeck_Phys.Rev.B_2018_p235452,Erpenbeck_J.Chem.Phys._2019_p191101,Erpenbeck_Phys.Rev.B_2020_p195421,Ke_J.Chem.Phys._2021_p234702} we have specifically studied the effect of multiple electronic states and multiple vibrational modes on the dissociation dynamics. 

The results obtained for a series of models of increasing complexity show the importance of multistate and multimode effects. For example, we found that vibronic coupling between $\pi^*$ and $\sigma^*$ states can enhance the dissociation rate at low bias voltages profoundly. This scenario is expected to be of importance for the rupture of bonds to heteroatoms in aromatic molecules, as has already been observed in the related though simpler processes of photoinduced dissociation dynamics and dissociative electron attachment in aromatic molecules such as pyrrole in the gas phase.\cite{Muendel_Phys.Rev.A_1985_p181,Modelli_J.Phys.Chem.A_2001_p5836--5841,Skalicky_Phys.Chem.Chem.Phys._2002_p3583--3590,Vallet_J.Chem.Phys._2005_p144307,Rescigno_Phys.Rev.Lett._2006_p213201,Chung_Phys.Chem.Chem.Phys._2007_p2075--2084,Oliveira_J.Chem.Phys._2010_p204301,Janeckova_Phys.Rev.Lett._2013_p213201,Slaughter_Phys.Chem.Chem.Phys._2020_p13893--13902,Dvorak_Phys.Rev.A_2022_p062821,ragesh2022distant}
Furthermore, in the high-bias resonant transport regime, an reaction pathway combining vibrational heating with a subsequent direct transition to the dissociative surface was found to play an important role.

The present investigation, in combination with previous studies of simpler models,\cite{Erpenbeck_Phys.Rev.B_2018_p235452,Erpenbeck_J.Chem.Phys._2019_p191101,Erpenbeck_Phys.Rev.B_2020_p195421,Ke_J.Chem.Phys._2021_p234702} provides a comprehensive analysis of the mechanisms of current-induced bond rupture prevailing in different parameter regimes.
It is of relevance also for STM-studies of current-induced reactions in molecules at metal surfaces.
Moreover, it can build the basis for the investigation of current-induced bond rupture in more complex systems and, in a broader context, of current-induced chemical reactions in general. 
This may also require a further advancement of the methodology. Possible directions include, but are not limited to, the semiclassical treatment of low-frequency vibrational modes where the quantum mechanical treatments are expensive,\cite{PhysRevB.107.115416,Preston_Phys.Rev.B_2020_p155415,Preston_J.Chem.Phys._2021_p114108} as well as the development of more optimized tensor network structures and more advanced time-propagation schemes.\cite{Wang_J.Phys.Chem.A_2013_p7431-7441,Shi_Phys.Rev.A_2006_p022320,Borrelli_J.Phys.Chem.B_2021_p5397--5407,Yang_Phys.Rev.B_2020_p094315,Dunnett_Phys.Rev.B_2021_p214302,Yan_J.Chem.Phys._2020_p204109,Yan_J.Chem.Phys._2021_p194104,Xu_JACSAu_2022_p335--340,Li_arXivpreprintarXiv:2208.10972_2022_p}

\section*{Acknowledgements}
The authors thank A. Erpenbeck and U. Peskin for helpful discussions. This work was supported by the German Research Foundation (DFG). Furthermore, the authors acknowledge support by the High Performance and Cloud Computing Group at the Zentrum für Datenverarbeitung of the University of Tübingen, the state of Baden-Württemberg through bwHPC and the German Research Foundation (DFG) through grant no INST 40/575-1 FUGG (JUSTUS 2 cluster) and INST 37/935-1 FUGG (BINAC cluster).

\section*{Supplementary Material}
See the supplementary material for the  details  of  (1) dissociation and population dynamics of the electronic states as well as the corresponding wave packet dynamics at $\Phi=1$ V and 3 V; (2) conducting properties for different models at different bias voltages; (3) the respective influence of the constant and coordinate-dependent terms in $\Delta(x_2)$ on the dissociation; (4) the analysis of the different forms of the diabatic coupling $\Delta(x_2)$ in the numerical performance of the HEOM+MPS/TT method.

\section*{Data Availability}
The data that support the findings of this study are available from the corresponding author upon reasonable request.

\footnotesize
%
\end{document}